# OCEP: An Ontology-Based Complex Event Processing Framework for Healthcare Decision Support in Big Data Analytics

Ritesh Chandra*, Sonali Agarwal, Shashi Shekhar Kumar, and Navjot Singh
Indian Institute of Information Technology Allahabad, India
*rsi2022001@iiita.ac.in, sonali@iiita.ac.in, rsi2020502@iiita.ac.in, navjot@iiita.ac.in*

**Abstract**

The exponential expansion of real-time data streams across multiple domains needs the development of effective event detection, correlation, and decision-making systems. However, classic Complex Event Processing (CEP) systems struggle with semantic heterogeneity, data interoperability, and knowledge driven event reasoning in Big Data environments. To solve these challenges, this research work presents an Ontology based Complex Event Processing (OCEP) framework, which utilizes semantic reasoning and Big Data Analytics to improve event driven decision support. The proposed OCEP architecture utilizes ontologies to support reasoning to event streams. It ensures compatibility with different data sources and lets you find the events based on the context. The Resource Description Framework (RDF) organizes event data, and SPARQL query enables rapid event reasoning and retrieval. The approach is implemented within the Hadoop environment, which consists of Hadoop Distributed File System (HDFS) for scalable storage and Apache Kafka for real-time CEP based event execution. We perform a real-time healthcare analysis and case study to validate the model, utilizing IoT sensor data for illness monitoring and emergency responses. This OCEP framework successfully integrates several event streams, leading to improved early disease detection and aiding doctors in decision-making. The result shows that OCEP predicts event detection with an accuracy of 85%. This research work utilizes an OCEP to solve the problems with semantic interoperability and correlation of complex events in Big Data analytics. The proposed architecture presents an intelligent, scalable and knowledge driven event processing framework for healthcare based decision support.

Table 1: Abbreviations

| Abbreviation | Full Form |
|---|---|
| RDF | Resource Description Framework |
| OWL | Ontology Web Language |
| API | Application Programming Interface |
| IoT | Internet of Things |
| OCEP | Ontology-Based Complex Event Processing |
| KRR | Knowledge Representation and Reasoning |
| SigMR | Signature-based MapReduce |
| H2RDF | Hadoop to RDF |
| HDT | Header, Dictionary, Triples |
| ETL | Extract, Transform, Load |
| BDC | Big Data Campaign |
| OBDA | Ontology-based Data Access |
| STARQL | Streaming and Temporal ontology Access with a Reasoning-based Query Language |
| BIGOWL4DQ | Big Data and Ontology Web Language for Data Quality |
| OUSAF | Ontology Usage Analysis Framework |
| EvCBR | Event-based Case-Based Reasoning |
| BIDMC | Beth Israel Deaconess Medical Center |
| PPG | Photoplethysmography |

## 1. Introduction

The IoT integrates smart devices with sensors and various identifiers to facilitate seamless interaction and data transfer. It transforms physical objects into smart systems that enhance traceability, visibility, and management in domains like forest fire [1], healthcare [2], and smart homes and smart cities [3].

Due to various structures, types, and domains, IoT devices which include sensors, gateways, actuators, and mobile phones generate huge amounts of real-time data, resulting in complex interoperability challenges. The interoperability issues encompass device, network, syntactic, semantic, and platform compatibility. Semantic interoperability is challenging as it refers to the data and semantic layers of IoT [4]. It ensures a unified comprehension of information from many data sources, facilitating the integration of diverse sensors and enhancing machine interpretation across several IoT applications. Subsequently, it improves by facilitating considerable correlations, rapid data integration, and knowledge based event reasoning [5][6].

Despite several explorations, traditional CEP systems have challenges related to semantic heterogeneity, data integration, and contextual reasoning inside dynamic IoT contexts. Rule-based methodologies depend on predefined patterns, limiting adaptability and scalability for complex event streams. Additionally, the absence of knowledge-based procedures inhibits efficient decision support in domains such as healthcare monitoring [7] [8].

An OCEP paradigm integrating semantic reasoning and extensive data analytics is proposed to address these challenges. It implements ontologies for semantic enhancement, RDF for structured representation, and SPARQL query for effective event reasoning. HDFS offers scalable storage, whereas Apache Kafka facilitates real-time data ingestion and processing.

The proposed OCEP framework is validated with a real-time healthcare use case that leverages IoT-based PPG sensors for early disease detection. By semantically integrating multiple source event streams, OCEP improves event detection and clinical decision support. The results demonstrate an overall 85% accuracy of the event detection, which outperforms traditional CEP systems. This research emphasizes the value of OCEP for an intelligent, scalable decision support in a Big Data healthcare environment.

Healthcare data analysis often includes comparing extracted measures to predefined thresholds. Symptoms are identified when a value is more or less than a given criteria. Early detection of the disease symptoms aids in the prediction of any major condition and so allows for their avoidance. The most important responsibility is to determine the accurate threshold. The accuracy of analysis depends heavily on the correctness of the thresholds used. A mathematical model with predefined parameters is used to achieve an accurate threshold.

The following are the main contributions of this manuscript which follows as:

1. Optimized Big Data Query Processing: Implements RDF-based structured data representation and utilizes SPARQL with MapReduce and HDFS for efficient, scalable, and parallelized query execution.
2. OCEP Framework: Leverages Hadoop and Apache Kafka for high-throughput, real-time data processing in dynamic event-driven analytics.
3. SSN Ontology-Driven Integration: Standardizes and integrates heterogeneous data sources, ensuring seamless interoperability in Big Data environments.
4. Ontology-Based Reasoning: Enhances real-time event recognition, knowledge extraction, and intelligent decision-making in complex data streams.

The organization of the manuscript is as follows: Section II provides background analysis and related work to identify the research gap in this domain. Section III outlines the materials and methodology used in this study. Section IV presents the experimental details along with the corresponding results. Section V concludes the manuscript and discusses the proposed framework.

## 2. Background Analysis and Related Works

To identify the research gap, it is essential to conduct a literature survey and demonstrate how the proposed solution addresses the shortcomings of existing research in this domain.

### 2.1 Ontology with Big Data Frameworks

Konys [10] proposed a framework for integrating Big Data, IoT, and the Semantic Web to create an enhanced platform for the emerging digital era. The integration was validated through real-life use cases, demonstrating its effectiveness in extracting valuable information from various sources.

Castro et al. [11] proposed an ontological reasoning-based approach to simplify data complexity. They employed semantic techniques for reasoning-driven data governance. To validate the proposed approach, a prototype was developed within a Big Data framework to demonstrate its effectiveness in reducing complexity.

Nural et al. [12] proposed a semantic technology-based approach for selecting appropriate models for data analysis. They utilized analytics-driven ontological models to enable semi-automated inferences. The SCALATION framework, which currently supports over more than thirty modeling techniques for predictive Big Data analytics, served as a testbed for evaluating the effectiveness of the proposed semantic technology.

González et al. [13] proposed BIGOWL, an ontology-based model for Big Data analytics. It was designed to identify components connected through data sources and facilitate visualization by considering parameters, constraints, and formats. To validate the approach, they conducted two case studies: first, real-world stream processing of traffic open data using Spark for route optimization in the urban environment of New York City; and second, data mining and classification of an academic dataset on local and cloud platforms. The analytics workflows generated by the BIG OWL semantic model were successfully validated and evaluated.

Redavid et al. [14] proposed a model-based approach for Big Data Analytics-as-a-Service (MBDAaaS). The proposed model, comprising declarative, procedural, and deployment sub-models, facilitates the selection of a deployable set of services based on user preferences, thereby shaping a BDC. Additionally, they employed an OWL ontology to address this challenge.

### 2.2 CEP with Healthcare

Dhillon et al. [15] proposed a CEP based model leveraging edge computing for remote healthcare monitoring using mobile devices that collect data from body attached sensors. The CEP predicts events based on the data collected through the hospital server. The performance of the model was evaluated using a synthetic workload, and provides insights into system scalability and workload parameters.

Rahmani et al. [16] proposed an event focused architecture for precise healthcare data analysis, containing three independent layers: service, event, context. The characteristics of the event layer, along with artificial intelligence for CEP methodologies to enhance reliability, reduce expenses, and elevate the quality of healthcare.

Yao et al. [17] proposed a novel CEP system for automating surgical events detection in hospitals using RFID technology. The proposed prototype was used for surgical management and evaluated for functionality and scalability. The performance evaluation shows significant improvements in patient safety and operational efficiency due to the framework's sense-and-response capability in detecting surgical events.

Mdhaffar et al. [18] introduced an innovative approach for forecasting heart failure by CEP using established thresholds based on patient body parameters. The threshold was determined by a statistical method based on historical data. The efficiency of the proposed method was evaluated utilizing several evaluation metrics.

Shashi et al. [19] introduced an innovative approach for the identification of CVD with a CEP model. The CEP regulations were formulated in accordance with standard WHO guidelines and fuzzy logic. Synthetic real-time data was used to evaluate the proposed method for detecting events and predicting CVD.

Naseri et al.[20] advocated using CEP for remote healthcare monitoring. They used a variety of rule-based reasoning techniques and verified the methodology against a big healthcare dataset. The trial findings showed that the PART rule-based system beat other algorithms, with an accuracy of 98.61 percent. Furthermore, the JRip rule-based learning method produced 16 rules with an accuracy of 97.32%.

### 2.3 Knowledge Driven Event Reasoning

Anicic et al. [21] suggested a novel method for combining knowledge-rich reasoning to discover links between event streams. They proposed a rule-based language for pattern matching with a precise syntax and declarative semantics.

Keeney et al. [22] presented a strategy to improve the expressiveness of knowledge-based middleware by proposing three temporal operators for event matching. They created a prototype that demonstrates temporal correlation between two fault-managed network scenarios. The prototype was then assessed and scaled utilizing broader knowledge-based networks.

Shirai et al. [23] proposed a novel approach for event prediction using the EvCBR model, which applies case-based reasoning to identify cause-effect event pairs within a Knowledge Graph (KG). Unlike traditional KG completion models, EvCBR enables inductive link prediction without prior training, overcoming common limitations. The model can integrate with portions of the Wikidata KG to query relevant entities, predict ongoing events, and visualize past event cases along with the reasoning paths used for making predictions.

Delgoshaei et al.[24] proposed an approach to demonstrate how artificial intelligence, semantic data models, and machine learning can transform intelligent buildings through advanced KRR. Using the Apache Jena API, RDF, OWL and SPARQL, the approach enables multi-domain knowledge integration for applications like energy auditing, fault detection, and building-to-grid integration. The proposed semantic infrastructure enhances building analytics, optimizing system performance across mechanical, lighting, shading, and security domains.

### 2.4 MapReduce-based SPARQL for RDF Data Analytics

Ahn et al. [25] proposed a novel approach called SigMR, a pruning method for multi-way join-based SPARQL query processing in MapReduce, aimed at addressing scalability challenges in querying large RDF datasets. By employing a signature encoding technique, SigMR significantly reduces the data size during query evaluation, enabling efficient processing within a single MapReduce job. Experimental results demonstrated faster query processing compared to existing MapReduce-based approaches.

Abdelaziz et al. [26] proposed a versatile framework that combines SPARQL queries with generic graph algorithms for complex RDF analytics. Built on vertex-centric frameworks like GraphLab and Pregel, it efficiently executes queries using an optimized SPARQL operator. Spartex scales to billion-edge datasets, outperforming state-of-the-art RDF engines by an order of magnitude for complex analytical tasks.

Ravindra [27] proposed an two-pronged approach for efficient RDF analytical query processing on MapReduce by reducing join-intensive tasks and intermediate data redundancy. It introduces a Nested TripleGroup Data Model and algebra to optimize graph pattern queries, minimizing MapReduce cycles. Integrated into Apache Pig, the approach achieves 45–60% performance gains over existing systems like Pig and Hive.

Mammo et al. [28] proposes  a novel approach of Presto-RDF, an architecture for efficient big RDF data processing using Facebook's Presto instead of traditional MapReduce systems like Hadoop. By translating SPARQL queries into SQL using a custom compiler and loading RDF data into HDFS, Presto-RDF demonstrated significantly higher performance compared to Apache Hive and the 4store RDF store across datasets ranging from 10 to 30 million triples.

Wang et al.[29] presented an efficient MapReduce-based method for distributed subgraph matching queries on large RDF graphs. The approach decomposes query graphs into star patterns, leveraging semantic and structural information as heuristics. Two optimizations RDF property filtering and delaying Cartesian product operations—enhance query performance. Experiments demonstrate that the method surpasses S2X and SHARD by an order of magnitude on average.

Papailiou et al. [30] proposed a novel methodology H2RDF, an efficient RDF store that handles complex SPARQL queries using distributed Merge and Sort-Merge joins with a multiple-index scheme over HBase indexes. A greedy planner with a cost model adapts query execution across single or multi-machine environments based on join complexity. The system, demonstrated via a web-based GUI, allows users to query datasets, monitor cluster status, and compare performance with other state-of-the-art RDF stores.

Papadaki et al.[31] presented surveys , approaches, systems, and tools for analytic queries over RDF-based Knowledge Graphs (KGs). It categorizes queries into domain-specific and quality-related, identifies five types of analytic approaches, and discusses aspects like efficiency and visualization. The survey aims to guide researchers and engineers in enhancing user-friendly analytics for Knowledge Graphs.

**2.5 Related Work**

Gimenez et al.[32] presented an approach to HDT-MR, a MapReduce-based algorithm for distributed RDF serialization into the HDT format. HDT efficiently compresses RDF data, enabling querying without decompression, thus addressing volume and velocity challenges. HDT-MR further improves scalability by distributing the serialization process, reducing resource consumption and processing time, while supporting larger datasets.

Banane et al.[33] introduced an approach of SPARQL2Hive, a SPARQL query processing system on MapReduce that enables efficient querying of large RDF graphs. It leverages Hive, a data warehouse system atop Hadoop, as an intermediate layer between SPARQL and MapReduce, ensuring compatibility with future Hadoop updates. Using ATL language for meta-model transformation, SPARQL2Hive outperforms other MapReduce-based SPARQL implementations.

Mallek et al. [34] explored the challenges Big Data poses to ETL processes in decision support systems, focusing on data volume, variety, and velocity. It reviews existing research and introduces

BRS-ETL, a novel approach designed to handle heterogeneous and streaming data, enhancing business intelligence workflows.

Sezer et al. [35] proposed a combined framework for integrating Big Data, IoT, and Semantic Web technologies to address limitations of traditional data processing systems. It outlines the system's components, their integration, and demonstrates the framework's effectiveness through a real-world use case, showcasing its potential for enhanced data processing and analytics.

Fathy et al.[36] addressed the challenge of accessing heterogeneous Big Data by proposing a unified semantic model for integration. It reviews existing approaches and introduces a semantic integration framework specifically for graph-based Big Data, enabling consistent access and maximizing the value of distributed datasets.

Schiff et al.[37] proposed an OBDA system using STARQL to query static, temporal, and streaming data stored in a SPARK cluster framework. It highlights the challenge of transforming high-level queries into backend-executable queries and the need for a homogeneous access interface. Experimental results demonstrate that achieving scalability with SPARK requires in-depth knowledge of data formats and organization.

González et al. [38] proposed a method of BIGOWL4DQ, an ontology-based framework for data quality measurement and assessment in AI workflows. It provides a machine-readable, interoperable vocabulary to integrate data quality analysis into Big Data pipelines. Validated through academic and real-world use cases, BIGOWL4DQ enhances reasoning capabilities and ensures data suitability for AI tasks.

Ashraf et al. [39] proposed an approach of OUSAF, a system for analyzing ontology usage in Big RDF Data. It offers a methodological approach to identify, analyze, and utilize ontology usage patterns, aiding ontology evolution, population, and deployment. The framework is validated using an e-business dataset, demonstrating its value for data publishers and ontology developers.

## 3. Materials and Methodology

This section presents a detailed description of the proposed architecture and its functionality. The model integrates various phases, including data collection through sensors, Apache Hadoop, Apache Kafka, and a CEP framework, along with scalable storage and ontology. This integration offers a comprehensive solution and depth analysis of the proposed approach.

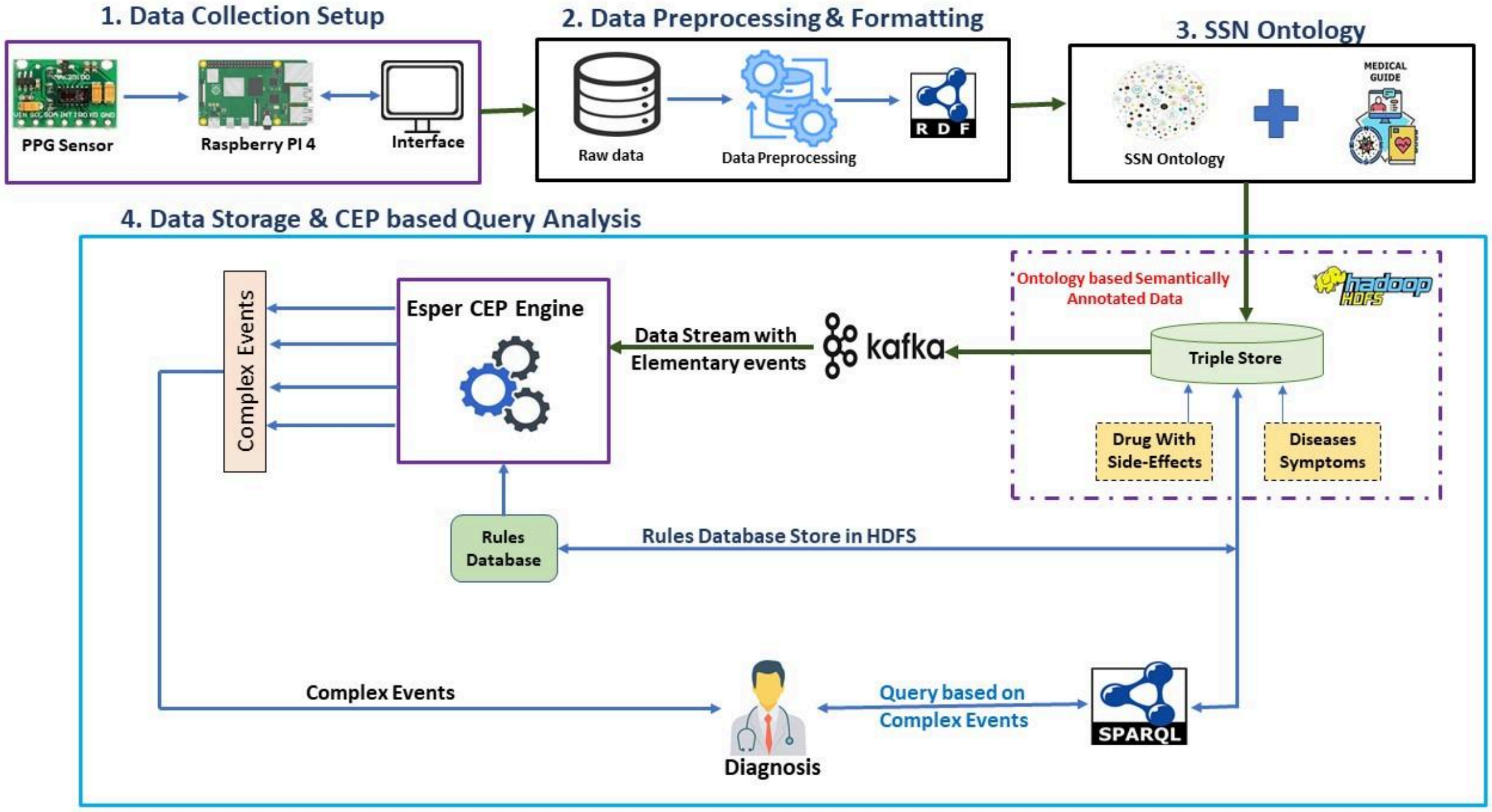

Figure 1. Proposed Architecture for OCEP

### 3.1 Data Collection Setup

We have used MAX30100 and Raspberry Pi 4 for data collection. The MAX30100 sensor collects four different physiological parameters: Heart Rate (HR), Pulse (PULSE), respiration rate and oxygen saturation ($SpO_2$) using an optical photoplethysmography (PPG) technique. The sensor emits infrared and red light into the skin, and a photodetector measures the amount of light reflected or absorbed by the blood. As blood volume changes with heartbeat, the intensity of reflected light varies, creating a pulse waveform from which heart rate is determined by analyzing the intervals of the PPG signal.

The pulse waveform derives from real-time PPG data, reflecting vascular health and circulation patterns. The $SpO_2$ level is computed by comparing the absorption ratios of red and IR light, as oxygenated and deoxygenated blood absorbs these wavelengths differently. The RESP is inferred by detecting low-frequency modulations in the PPG waveform caused by thoracic pressure changes during breathing. To estimate respiratory cycles, these fluctuations are analyzed using baseline wander analysis, wavelet transforms, or Fast Fourier Transform (FFT) [40]. The setup with a pin diagram for data collection is shown in Figure 2. While Figure 1 shows the proposed architecture of OCEP.

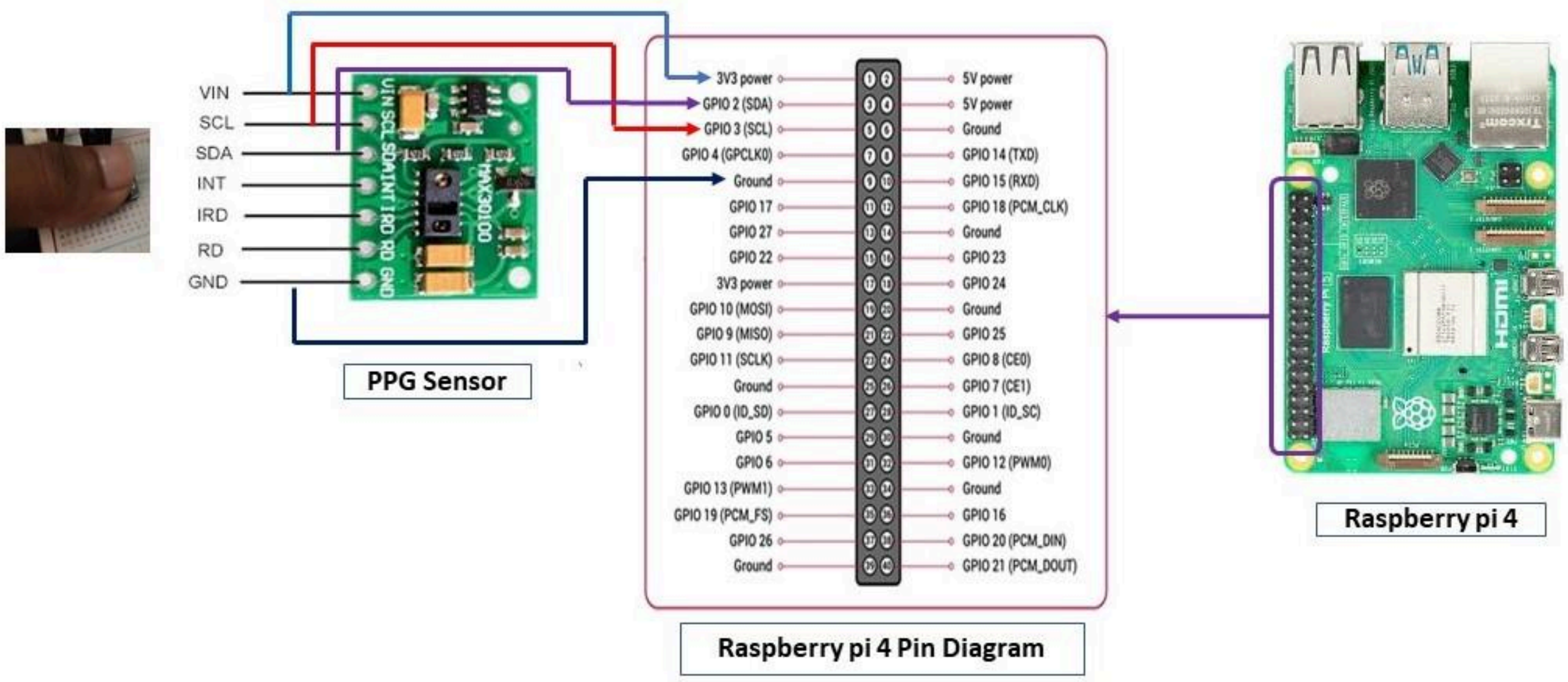

Figure 2. Data collection setup using PPG and Raspberry Pi

### 3.2 Data Preprocessing

The collected data is in CSV format, comprising PPG recordings from 28 individuals and 12 minutes of each individual record. However, this sample size is insufficient for robust analysis. We incorporated open-source data from the BIDMC PPG dataset[1], which contains records from 53 patients, each featuring 8 minutes of recorded data. This dataset provides PPG signals and labeled disease information, enabling a more comprehensive analysis of how PPG patterns correlate with different medical conditions.

Figure 3 shows the dataset's transformation before and after preprocessing. The raw data exhibits irregularities, such as abrupt spikes, missing values, and outliers, primarily due to sensor noise or data artifacts. Post-preprocessing, the data shows cleaned and normalized trends for HR, RR, and SpO2. Key preprocessing steps included handling missing values, correcting outliers, and applying normalization. These steps effectively eliminated inconsistencies, smoothed spikes, and scaled values, resulting in a more reliable and representative dataset suitable for analysis. The physiological parameters are as follows:

- RR: Measured in breaths per minute, derived from the impedance signal.
- HR: Measured in beats per minute, obtained from the ECG.
- PR: Measured in beats per minute, derived from the PPG.
- SpO2: Recorded as a percentage.

[1] https://physionet.org/content/bidmc/1.0.0/

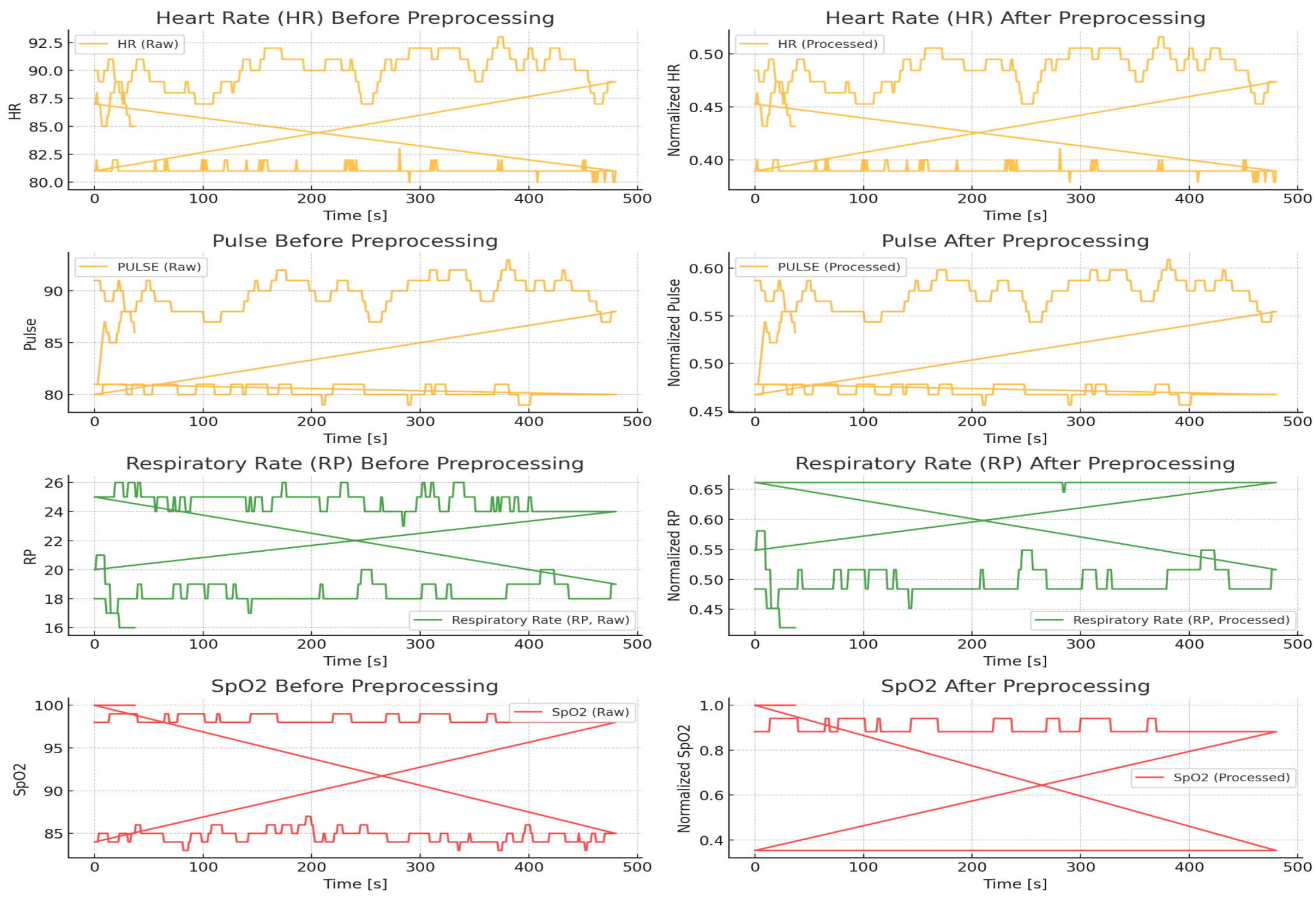

Figure 3. Dataset before Preprocessing and after Preprocessing

After preprocessing, we convert this dataset into RDF using RDFlib in python by mapping each patient's PPG dataset into RDF triples. The script reads the CSV file using pandas, assigns a unique URI for each patient, and defines RDF properties for HR, PULSE, RESP, and $SpO_2$ using a custom namespace (http://Healthcare.org/ppg/) [41]. The RDF graph is then populated with these triples and serialized into Turtle format (.rdf) for interoperability with semantic web applications. This structured representation enables efficient querying, reasoning, and integration with linked healthcare datasets. Figure 4 shows the RDF conversion using the RDF library.

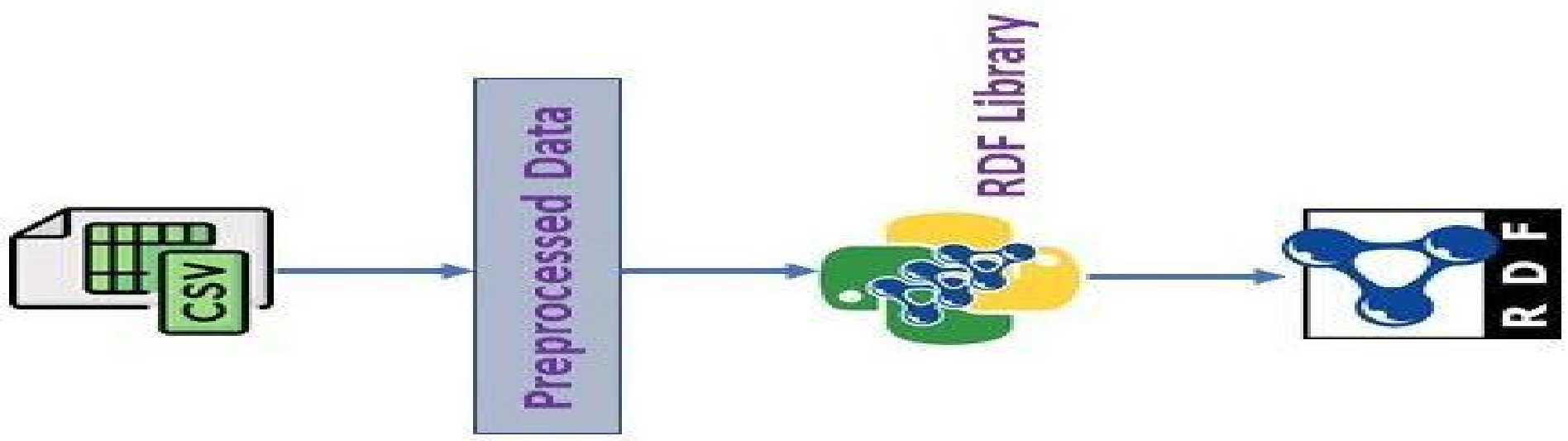

Figure 4. CSV data to RDF conversion using RDF Library

Table 2: Conversion of CSV to RDF data

| **Algorithm 1: Convert_PPG_CSV_to_RDF_data** |
|---|
| Input: CSV file containing PPG sensor data<br>Output: RDF graph serialized in Turtle format<br><br>Step 1. Initialize RDF Library:<br>a. Import rdflib<br>b. Create an RDF Graph (g)<br>c. Define Namespace ( "http://healthcare.org/ppg")<br>d. Import RDF standard namespaces (RDF, XSD)<br><br>Step 2.. Load CSV_file into DataFrame (df)<br><br>Step 3. Define RDF Type and Properties:<br>- SSN:PPGData<br>- SSN:hasTime<br>- SSN:hasHR<br>- SSN:hasPULSE<br>- SSN:hasRESP<br>- SSN:hasSpO2<br><br>Step 4. For each row in df:<br>a. Create a unique URI for each timestamp (SSN:Time_Timestamp)<br>b. Add RDF Triples:<br>- (Time_URI, rdf:type, ex:PPGData)<br>- (Time_URI, SSN:hasTime, Literal(Time, datatype=XSD.integer))<br>- (Time_URI, SSN:hasHR, Literal(HR, datatype=XSD.integer))<br>- (Time_URI, SSN:hasPULSE, Literal(PULSE, datatype=XSD.integer))<br>- (Time_URI, SSN:hasRESP, Literal(RESP, datatype=XSD.integer))<br>- (Time_URI, SSN:hasSpO2, Literal(SpO2, datatype=XSD.integer))<br><br>Step 5. Serialize Graph in Turtle format<br>Step 6. Save as "ppg_data.ttl"<br>Step 7. End Algorithm |

### 3.3 Ontology Development

Semantic interoperability is crucial for seamless communication among IoT devices from various vendors, especially in healthcare, where diverse devices generate heterogeneous data. To address the challenge, a SSN and sensors, Observation, Sample, and Actuator (SOSA) ontology is deployed, enabling IoT devices to semantically annotate and describe the data in a structured manner using OWL. SSN provides a foundational framework for representing sensor networks and observations, while SSNO extends SSN with enhanced capabilities for sensor and actuator representation, spatial and temporal reasoning, and improved interoperability. SOSA, a lightweight ontology, improves SSN by focusing on observations, sensors, sampling, and actuators, hence enabling IoT based applications and large-scale data integration. These ontologies use OWL [1] to enable automated reasoning and machine readable data, to ensure meaningful interpretation by easy data sharing across heterogeneous IoT devices, as shown in Figure 5. Table 2 shows the conversion of PPG CSV to RDF.

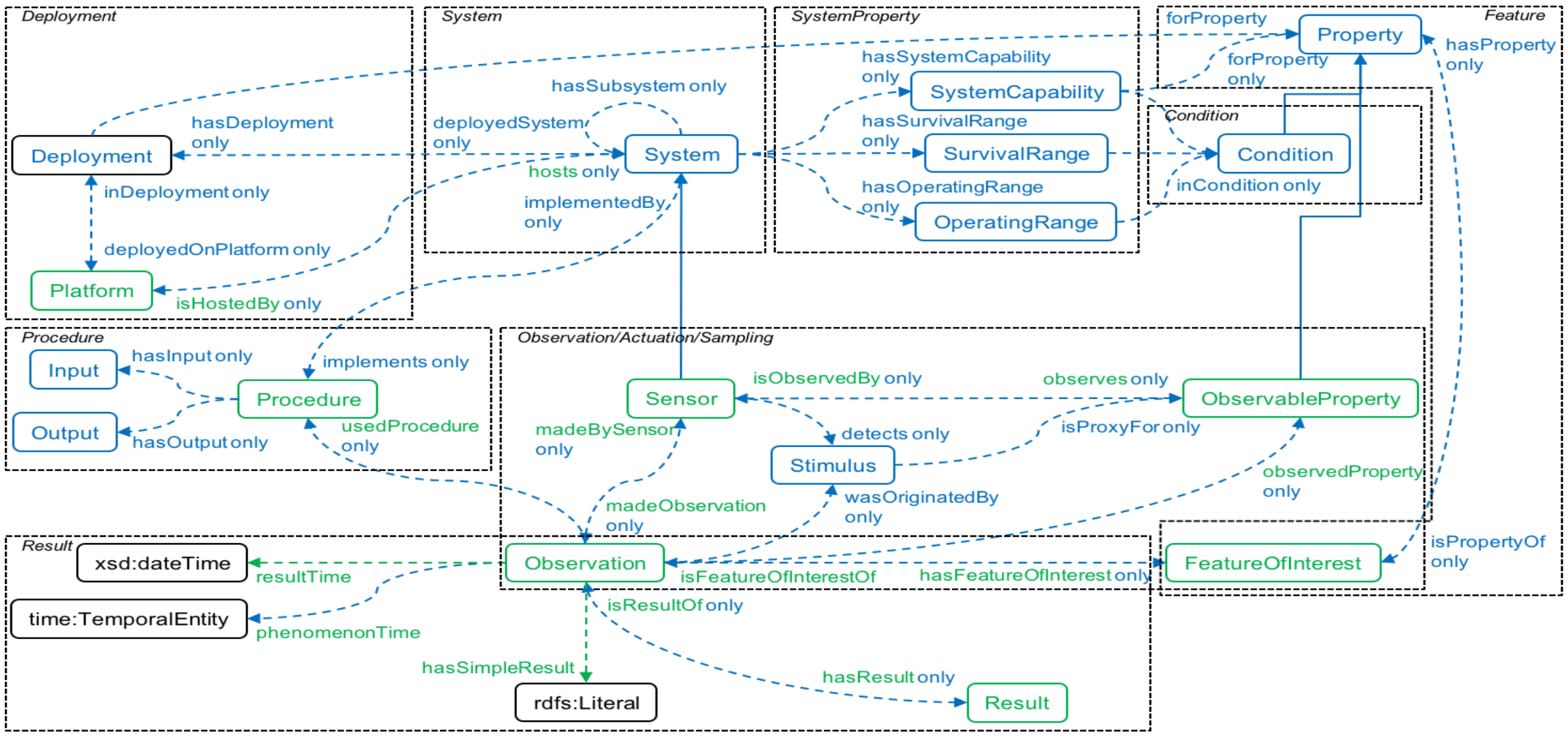

Figure 5. Overview of the SSN[2] Classes and Properties

The SSN ontology completely includes all aspects of PPG-based disorders, in accordance with medical norms for accurate disease representation and diagnosis. Merging RDF data with the SSN ontology allows for organized, semantically rich sensor data representation, assuring interoperability, reasoning, and integration across IoT devices. By aligning RDF triples with SSN classes and attributes, sensor data becomes self-descriptive, machine-readable, and queryable, allowing for frictionless data transmission, advanced semantic reasoning, and better decision-making in healthcare applications.

### 3.3.1 Core Classes of Ontology

The SSN and SOSA ontologies provide core classes essential for structured sensor data representation and interoperability. Sensor (SSN/SOSA) represents devices performing observations, while Observation captures measured phenomena. Feature of Interest denotes real-world entities like heart rate, and Observed Property defines characteristics such as blood oxygen levels. Stimulus triggers sensor responses, generating Results as data outputs. Procedure standardizes measurement methods, while the Platform hosts sensors, and Deployment defines their operational context [1]. The system represents sensor networks. SOSA adds an actuator for devices that perform actions, sampling for data extraction, a sample for subsets of observed data, and an ultimate feature of interest for broader contextual references, as shown in Figure 6. These classes ensure semantic interoperability, reasoning, and efficient data exchange in IoT-based healthcare applications.

---

[2] https://www.w3.org/TR/vocab-ssn/

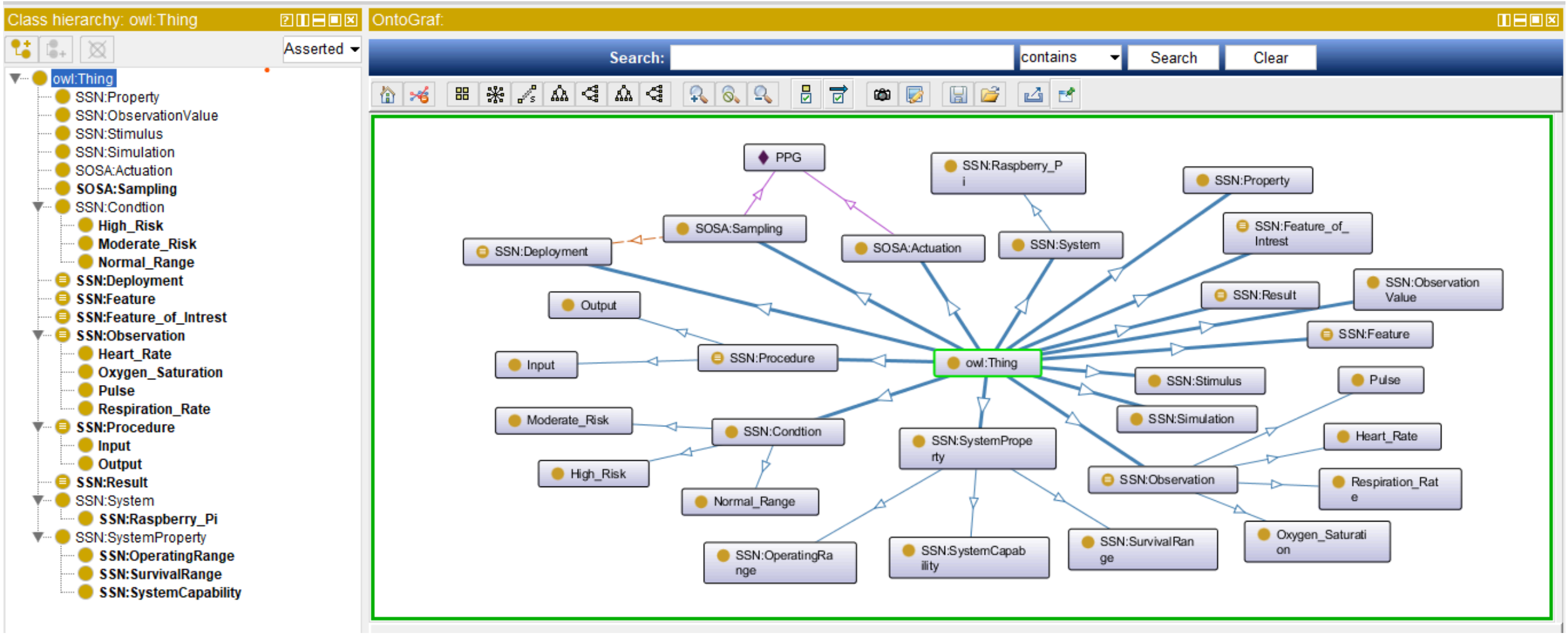

Figure 6: Core Classes of SSN and SOSA Ontology Created by Protege with OntoGraf

### 3.3.2 PPG Data-Based Disease Prediction Parameters Based on Medical Guidelines

The PPG Data-Based Disease Prediction Parameters table categorizes key physiological metrics into Normal, Moderate Risk, and High Risk based on medical guidelines shown in Table 1. Parameters like HR, $SpO_2$, PWV, and HRV help assess cardiovascular and respiratory health, aiding in early detection of hypertension, Tachycardias, COPD, CAD, and hypoxia. This structured approach enhances preventive care, real-time monitoring, and personalized health management. Symptoms and medicine will also be recommended by the ontology or retrieved by a SPARQL query. These parameters are also validated based on the 53 patient datasets based on this disease, which is correct because each patient type of disease is given in the dataset. Table 3 shows parameters based on medical guidelines

Table 3. Parameters based on Medical Guidelines

| Parameter | Normal Range | Moderate Risk | High Risk |
|---|---|---|---|
| Gender | Typically, lower cardiovascular risk for females | Males at moderate risk post-45 | Higher risk for males after 60 |
| Heart Rate (HR)[3] | 60-100 BPM | 100-120 BPM | Above 120 BPM |
| Blood Oxygen Saturation (SpO2)[4] | 95-100% | 90-94% | Below 90% |
| Pulse Wave Velocity (PWV)[5] | 5-9 m/s | 9-12 m/s | Above 12 m/s |
| Pulse Rate Variability[6] (PRV) | Above 50 ms | 30-50 ms | Below 30 ms |
| Respiration Rate (RR)[7] | 12-20 breaths/min | 20-24 breaths/min | Above 24 breaths/min |
| Systolic Blood Pressure (SBP)[8] | 90-120 mmHg | 120-140 mmHg | Above 140 mmHg |
| Diastolic Blood Pressure (DBP)[9] | 60-80 mmHg | 80-90 mmHg | Above 90 mmHg |
| Heart Rate Variability (HRV)[10] | Above 50 ms | 30-50 ms | Below 30 ms |
| Perfusion Index (PI)[11] | Above 2% | 0.5-2% | Below 0.5% |

### 3.3.3 Rule Development for Disease prediction using Mathematical model

Mathematical models are crucial in dynamically adjusting thresholds based on context parameter fluctuations. By utilizing these models, designers can establish tolerance levels that align with specific application needs. These models are formulated using a set of mathematical functions, which are applied to modify and refine thresholds as required. Below, we describe the implemented mathematical models used for threshold adaptation.

[3] https://pmc.ncbi.nlm.nih.gov/articles/PMC6592896/
[4] https://www.medicalnewstoday.com/articles/321044
[5] https://pmc.ncbi.nlm.nih.gov/articles/PMC4609308/
[6] https://pmc.ncbi.nlm.nih.gov/articles/PMC5624990/
[7] https://www.verywellhealth.com/what-is-a-normal-respiratory-rate-2248932
[8] https://pmc.ncbi.nlm.nih.gov/articles/PMC4573449/
[9] https://www.health.harvard.edu/heart-health/a-look-at-diastolic-blood-pressure
[10] https://www.rupahealth.com/post/what-is-heart-rate-variabilit
[11] https://www.medrxiv.org/content/10.1101/2022.10.19.22281282v1.full.pdf

**1) Constant Mathematical Model:** A constant function is a mathematical function that maintains the same output value regardless of the input.

**2) Step Function Mathematical Model:** A step function is a discontinuous function composed of multiple constant segments, where each segment is defined over a specific interval within the function's domain. Furthermore, a step function is defined in an interval [a; b], which can be divided into a set of reals such as $a = a_0 < a_1 < a_2 \ldots < a_n = b$, so as the restriction in each sub interval [$a_k$; $a_{k+1}$ ] is a constant.

**3) Average and Level of Confidence Mathematical Model:** A straightforward technique for updating thresholds that adds the average of $p$ previous data values to the confidence in the average's level.

**4) Moving Average (MA):** Numerous modifications and applications have been suggested by scholars during its evolution [42]. The Simple Moving Average (SMA) is a fundamental kind of moving average. The mathematical formula is give as following equation

$$SMA(t) = \sum_{i=1}^{p} value\ i/p \qquad (1)$$

where value i represents the i-th value of the parameter and p represents the number of previous context parameter values.

**5) An enhancement of the SMA is the Weighted Moving Average (WMA)**[12]. Newer values are given more weight than historic ones. Formally, the following formula defines the WMA model.

$$Weighted\ MA(t) = \sum_{i=1}^{p} (p - i)\ value\ i / \sum_{i=1}^{p} (p - i) \qquad (2)$$

where p is the number of preceding context parameters and value *i* is the parameter's *i-th* value.

**6) Additionally, we take into account the Exponentially Weighted Moving Average (EWMA)**, which is a type of WMA. A weighting factor is assigned to each context parameter value in the time series by the EWMA model, a weighted moving average. The formula is as follows.

$$ExpoWMA(t) = \alpha . value(t) + (1 - \alpha) . ExpoWMA(t-1) \qquad (3)$$

- Where ExpoWeightedM A(0) is a constant.
- Value $t$ is the observation that at time $t$.
- p is the number of observations to be monitored.
- $0 < \alpha < 1$ is a constant, it represents a smoothing factor between 0 and 1. Commonly, α is calculated using the formula given below.

12 https://www.investopedia.com/ask/answers/071414/whats-difference-between-moving-average-and-weighted-moving-average.asp

$$\alpha = 2/n+1 \quad (4)$$

This mathematical model determines threshold values based on predefined disease permissible limits, ensuring they are within the range of actual values. The model enhances accuracy in detecting disease and making efficient decisions by correlating disease-specific parameters with calculated thresholds.

**Rule 1:** from Heart_Rate [heartRate < heartRate_threshold (100 BPM)] select heartRate, patientId, insert into (Less chances of Tachycardia);

**Rule 2:** from Heart_Rate [heartRate > heartRate_threshold (100 BPM)] select heartRate, patientId, insert into (Moderate chances of Tachycardia);

**Rule 3:** from Heart_Rate [heartRate > heartRate_threshold (120 BPM)] select heartRate, patientId, insert into (Tachycardia);

RULE 1, 2 and 3 determines the identification of cases where the heart rate exceeds the predefined threshold. Detecting tachycardia is vital for predicting potential risk of various diseases.

### 3.4 Ontology based Semantically Annotated Data Store in HDFS

An ontology-based semantically annotated data store in HDFS supports scalable, structured, and interoperable data management for large-scale applications. RDF data is stored in a triple format (subject-predicate-object) and integrated with standard ontologies like SSN and SOSA, ensuring semantic consistency and interoperability. Subsequently, a drug side effect dataset and a symptom dataset for various diseases were transformed into RDF format and merged into the semantically detailed data store.

#### 3.4.1 SPARQL-based querying using MapReduce in HDFS

SPARQL based querying using MapReduce in HDFS enables efficient distributed processing of large-scale RDF datasets. The RDF data is stored in HDFS formats like N-Triples or Turtle, ensuring sufficient storage. The MapReduce framework processes SPARQL queries by parallelizing triple pattern matching. The Mapper reads RDF triples, extracts subject-predicate-object relationships, and produces key-value pairs, whereas the Reducer aggregates data, applies SPARQL filters, and returns query results. For example, a SPARQL query identifying patients with heart rates more than 100 beats per minute and Tachycardia would be processed and distributed using Mappers to parse triples and Reducers to filter criteria. Integrating SPARQL with MapReduce guarantees high-performance semantic querying, reasoning, and knowledge extraction across large RDF datasets in parallel on a distributed platform [25]. SPARQL query for patients having a heart rate more than 120 BPM and Tachycardia is shown in Table 4. It defines namespaces and standard data types through the use of various prefixes.

Table 4 Sample SPARQL query

**SPARQL query**

```
PREFIX rdf: <http://www.w3.org/1999/02/22-rdf-syntax-ns#>
PREFIX rdfs: <http://www.w3.org/2000/01/rdf-schema#>
PREFIX xsd: <http://www.w3.org/2001/XMLSchema#>

SELECT ?patient ?sideEffect
WHERE {
   ?patient SSN:hasHeartRate ?hr .
   ?patient SSN:hasCondition "Tachycardia" .
   ?patient SSN: takes medication?medication .
   ?medication Drug:hasSideEffect ?sideEffect .

   FILTER(?hr > 120)
}
```

The SparqlMapper.java class reads each line, extracts the subject, predicate, and object, and returns key-value pairs, where the key is the subject (Patient ID) and the value is the predicate-object pair (for example, hasHeartRate 110). This ensures that all relevant data for each patient is grouped together in the Reducer phase, when filtering conditions like Heart Rate > 120 and Condition = Tachycardia are used to execute queries. Table 5 shows mapper code for SPARQL query.

Table 5. Mapper code for SPARQL query in Table 4

**Mapper Code (SparqlMapper.java)**

```
public class SparqlMapper extends Mapper<LongWritable, Text, Text, Text> {
   private static final String PATIENT_PREFIX = "SSN:Patient";
   private static final String HEART_RATE_PREFIX = "SSN:hasHeartRate";
   private static final String CONDITION_PREFIX = "SSN:hasCondition";
   private static final String MEDICATION_PREFIX = "SSN:takesMedication";
   private static final String SIDE_EFFECT_PREFIX = "Drug:hasSideEffect";
   private static final Pattern TRIPLE_PATTERN = Pattern.compile("<(.+?)> <(.+?)> \"?(.+?)\"?\\.?");
   @Override
   protected void map(LongWritable key, Text value, Context context) throws IOException,
InterruptedException {
```

```
        String line = value.toString().trim();
        Matcher matcher = TRIPLE_PATTERN.matcher(line);
        String patient = null;
        double heartRate = 0;
        String condition = null;
        String medication = null;
        String sideEffect = null;
        while (matcher.find()) {
          String subject = matcher.group(1);
          String predicate = matcher.group(2);
          String object = matcher.group(3);

          if (predicate.endsWith(HEART_RATE_PREFIX)) {
            try {
              heartRate = Double.parseDouble(object);
            } catch (NumberFormatException e) {
              continue;}
          } else if (predicate.endsWith(CONDITION_PREFIX) && object.equals("Tachycardia")) {
            condition = object;
          } else if (predicate.endsWith(MEDICATION_PREFIX)) {
            medication = object;
          } else if (predicate.endsWith(SIDE_EFFECT_PREFIX)) {
            sideEffect = object;
          }
          if (patient == null && subject.startsWith(PATIENT_PREFIX)) {
            patient = subject;
          }
        }
        // Emit only if heart rate > 120 and condition is Tachycardia
        if (patient != null && heartRate > 120 && condition != null && medication != null &&
sideEffect != null) {
          context.write(new Text(patient), new Text(sideEffect));
        }
```

```
    }
}
```

The SparqlReducer.java class collects all predicate-object pairs for each patient from the Mapper phase and stores them in a HashMap to maintain track of data such as heart rate and medical status. It then performs SPARQL-like filtering, determining whether a patient has a heart rate greater than 100 BPM and Tachycardia. If both conditions are met, the reducer returns the patient ID as an output. This ensures that the query results only contain patients that meet the provided criteria, thereby simulating SPARQL query execution in a distributed Hadoop environment. Table 6 shows the reducer code for SPARQL query.

Table 6. Reducer code for SPARQL query in Table 4

**Reducer Code (SparqlReducer.java)**

```
public class SparqlReducer extends Reducer<Text, Text, Text, Text> {

        @Override
        protected void reduce(Text key, Iterable<Text> values, Context context)
        throws IOException, InterruptedException {

        StringBuilder sideEffectsList = new StringBuilder();

        // Aggregate all side effects for each patient
        for (Text value : values) {
        if (sideEffectsList.length() > 0) {
                sideEffectsList.append(", ");
        }
        sideEffectsList.append(value.toString());
        }

        // Write the result: (Patient -> List of Side Effects)
        context.write(key, new Text(sideEffectsList.toString()));
        }}
```

The SparqlDriver.java class is an entry point for the hadoop MapReduce job, which configures and executes the whole data processing pipeline. It first configures the hadoop task by defining the Mapper (SparqlMapper.java), Reducer (SparqlReducer.java), and output key-value types (Text, Text). The driver establishes the input and output channels for the RDF data in HDFS, ensuring that it flows properly between map and reduce phases. At last, it sends the job to the hadoop cluster and waits for completion before quitting with a success or failure condition. This class successfully incorporates SPARQL-like query execution into a scalable, distributed MapReduce workflow. Figure 7 explains how the SPARQL query is executed using MapReduce in HDFS on the RDF dataset. Table 7 shows the driver code for SPARQL query.

Table 7. Driver code for SPARQL query in Table 4

**Driver Code (SparqlDriver.java)**

```
public class SparqlDriver {
        public static void main(String[] args) throws Exception {
        if (args.length != 2) {
        System.err.println("Usage: SparqlDriver <input path> <output path>");
        System.exit(-1);
        }

        Configuration conf = new Configuration();
        Job job = Job.getInstance(conf, "SPARQL Processing");

    job.setJarByClass(SparqlDriver.class);
    job.setMapperClass(SparqlMapper.class);
    job.setReducerClass(SparqlReducer.class);

        job.setOutputKeyClass(Text.class);
    job.setOutputValueClass(Text.class);

        FileInputFormat.addInputPath(job, new Path(args[0]));
    FileOutputFormat.setOutputPath(job, new Path(args[1]));

    System.exit(job.waitForCompletion(true) ? 0 : 1);
        }
}
```

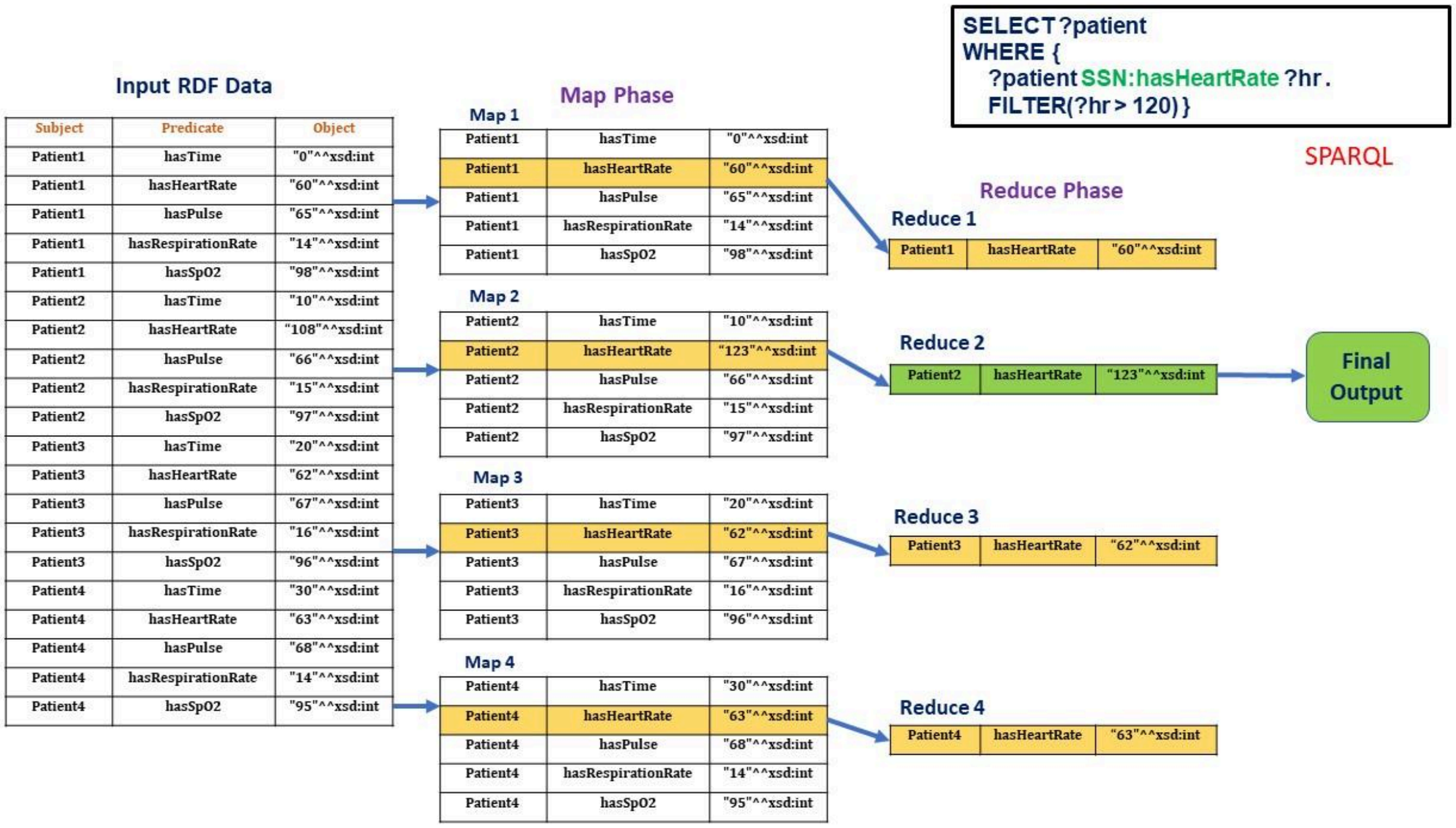

Figure 7: Steps involved in executing a SPARQL query using MapReduce in HDFS

### 3.4.2 SPARQL Query for Medication and Side Effects

The combination of two critical datasets, the Drug Dataset and the Disease Dataset, enables enhanced Big Data research in healthcare. The Drug Dataset, which has 111,079 MedDRA-validated items, provides thorough information on drugs and their adverse effects by using semantic data analytics to discover hidden patterns and relationships. Similarly, the Disease Dataset obtained from MedExpert has 81 data points describing disorders, recommended remedies, and potential side effects. Both datasets are transformed to RDF format and stored in the HDFS, where SPARQL searches give real-time information about pharmaceutical efficacy, safety profiles, and personalized treatment regimens. The architecture enables dynamic sickness classification using the most recent clinical recommendations overview shown in Figure 8.

Furthermore, IoT devices collect real-time symptoms from users via a variety of sensors, ensuring accurate and timely data collection. This data is used by the framework to deliver personalized prescription recommendations, food plans, and insights into potential side effects, thereby improving precision healthcare and optimal treatment approaches. Furthermore, query-based data analytics provide fast filtering, correlation, and prediction of health issues, allowing healthcare practitioners to extract meaningful insights from massive RDF datasets, hence enhancing clinical decision-making and patient outcomes.

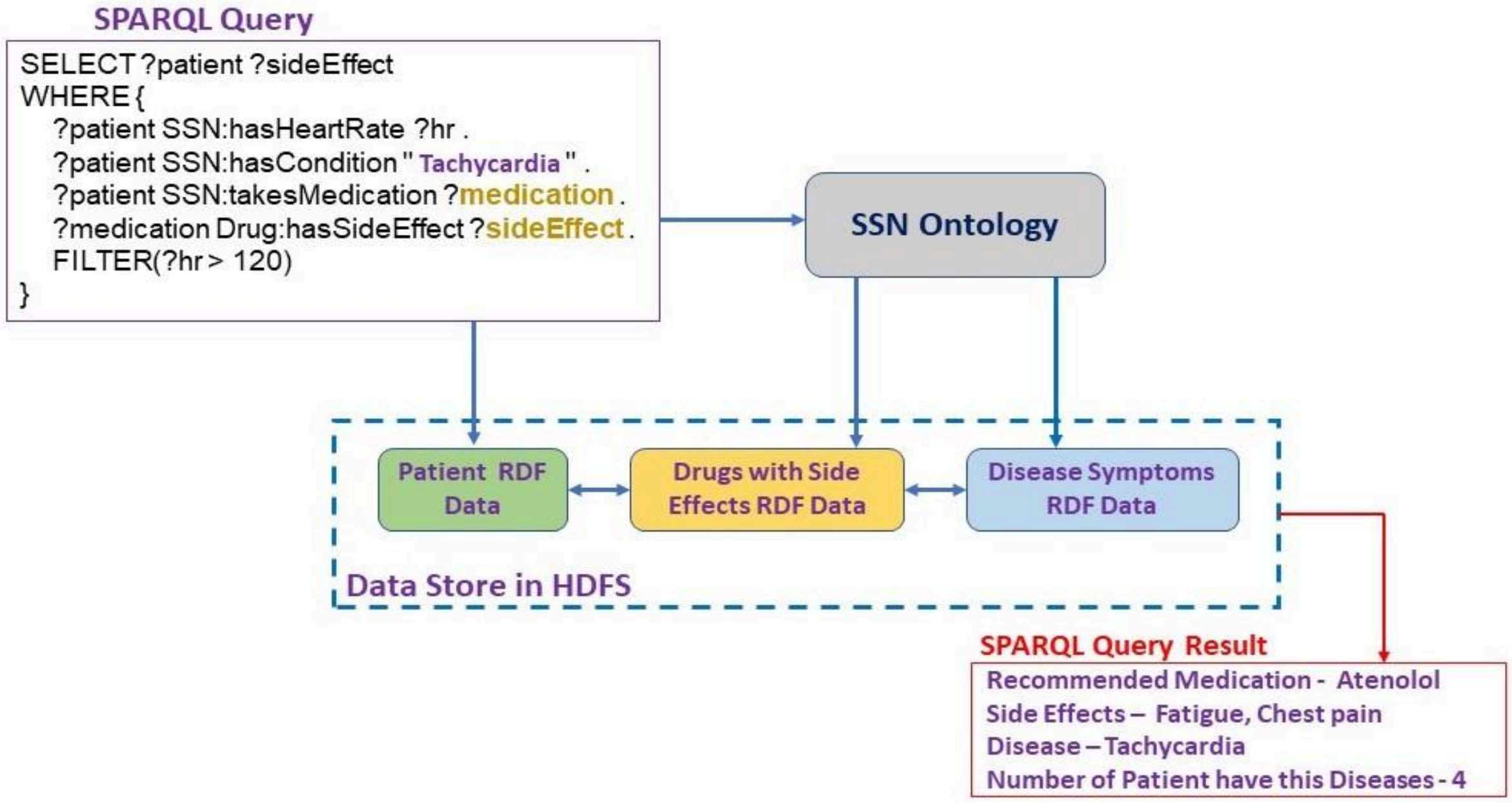

Figure 8: SPARQL query for medication and side effects

This SPARQL query retrieves 9 disease-drug pairs by selecting entities classified as diseases (SSN:Disease) and drugs (SSN:Drug), then filtering those where a drug treats a disease (SSN:treatedBy). Limiting the results is useful for data visualization, healthcare analytics, and drug recommendation systems, enabling efficient extraction of medical knowledge from RDF datasets shown in Table 8. Figure 9 shows the result for disease and drugs of SPARQL query.

Table 8: SPARQL Query

```
SELECT ?disease ?drug
WHERE { ?disease rdf:type SSN:Disease .
 ?drug rdf:type SSN:Drug .
?disease SSN:treatedBy ?drug .
} LIMIT 9
```

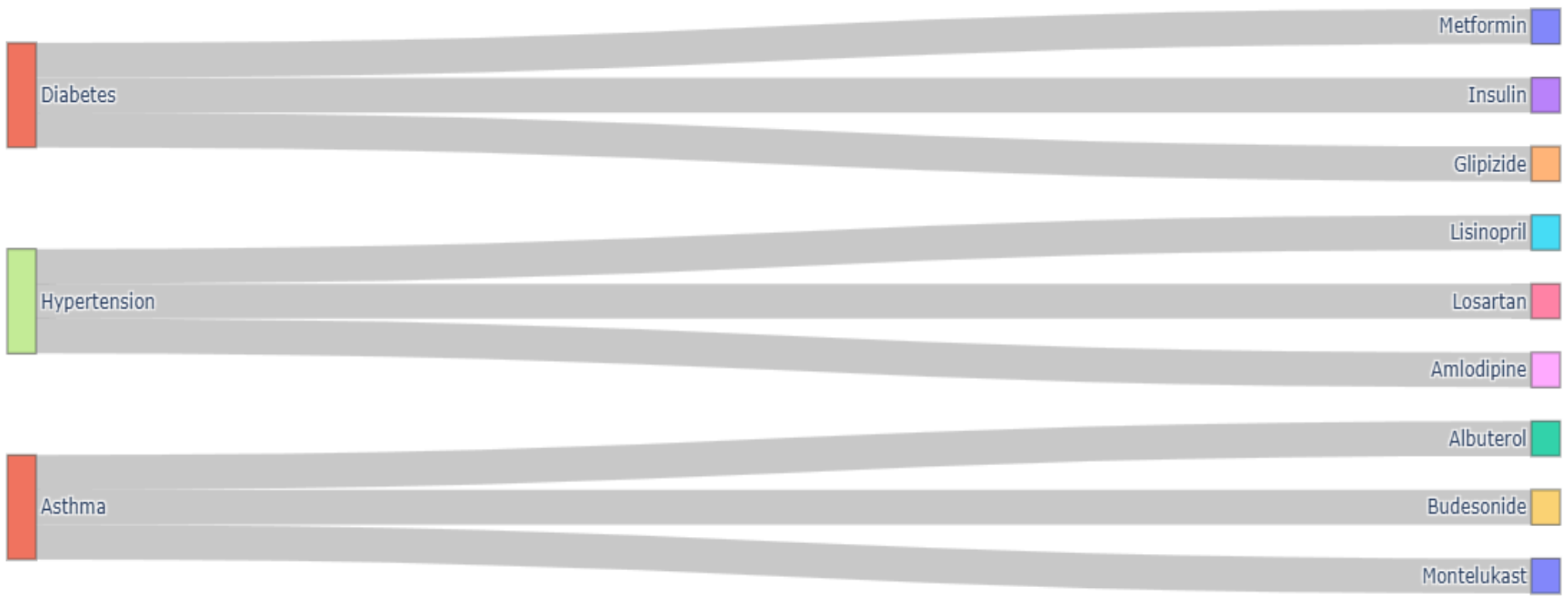

Figure 9. Result of disease and drugs of SPARQL query of table 6

### 3.5 HDFS stored RDF data to Kafka Preprocessing

Storing RDF data in HDFS provides a scalable and distributed framework for managing large-scale semantic datasets, but real-time processing and streaming analytics require integration with Apache Kafka. This is achieved using the Kafka Connect HDFS Sink Connector, which efficiently streams RDF triples from Kafka topics into HDFS in N-Triples format, ensuring exactly once delivery, partitioning, and fault tolerance. The RDF event flow begins with a Kafka producer that ingests RDF data via the publish-subscribe module, transmitting it to multiple brokers (A, B, C) for event replication and scheduling to mitigate potential data loss from broker failures shown in Figure 10. The Kafka consumer then parses and processes RDF streams, dividing them into structured events for further analysis. This integration enables high-throughput RDF data pipelines, allowing real-time querying and facilitating efficient knowledge graph updates and semantic reasoning at scale.

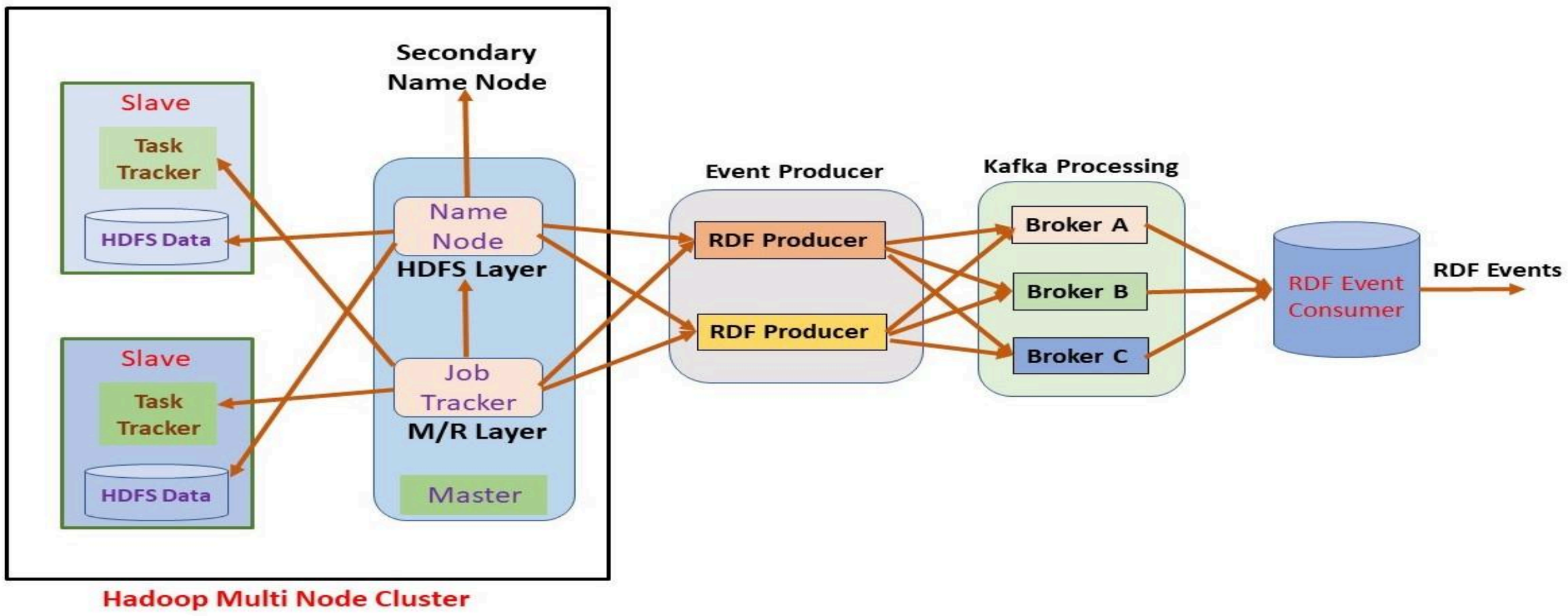

Figure 10. HDFS to RDF stream processing through Kafka

### 3.6 CEP operation using Esper

CEP with Esper for PPG Data allows for real-time PPG signal analysis, ensuring accurate cardiovascular health monitoring. Esper CEP[13] utilizes Apache Kafka to stream PPG data from IoT-enabled wearables and medical devices, continually processing heart rate, PWV, $SpO_2$, and HRV events shown in Figure 11. Predefined Esper EPL (Event Processing Language) rules detect anomalies such as tachycardia (HR > 100 BPM), Tachycardia patterns, or hypoxia ($SpO_2$ < 90%), prompting early action.

Esper utilizes event correlation to identify trends, such as increasing heart rate variability and abnormal pulse waves, which could suggest a cardiovascular risk. The analyzed events are then saved in HDFS in RDF format, allowing for further analysis using SPARQL queries for additional medicine and other recommendations.

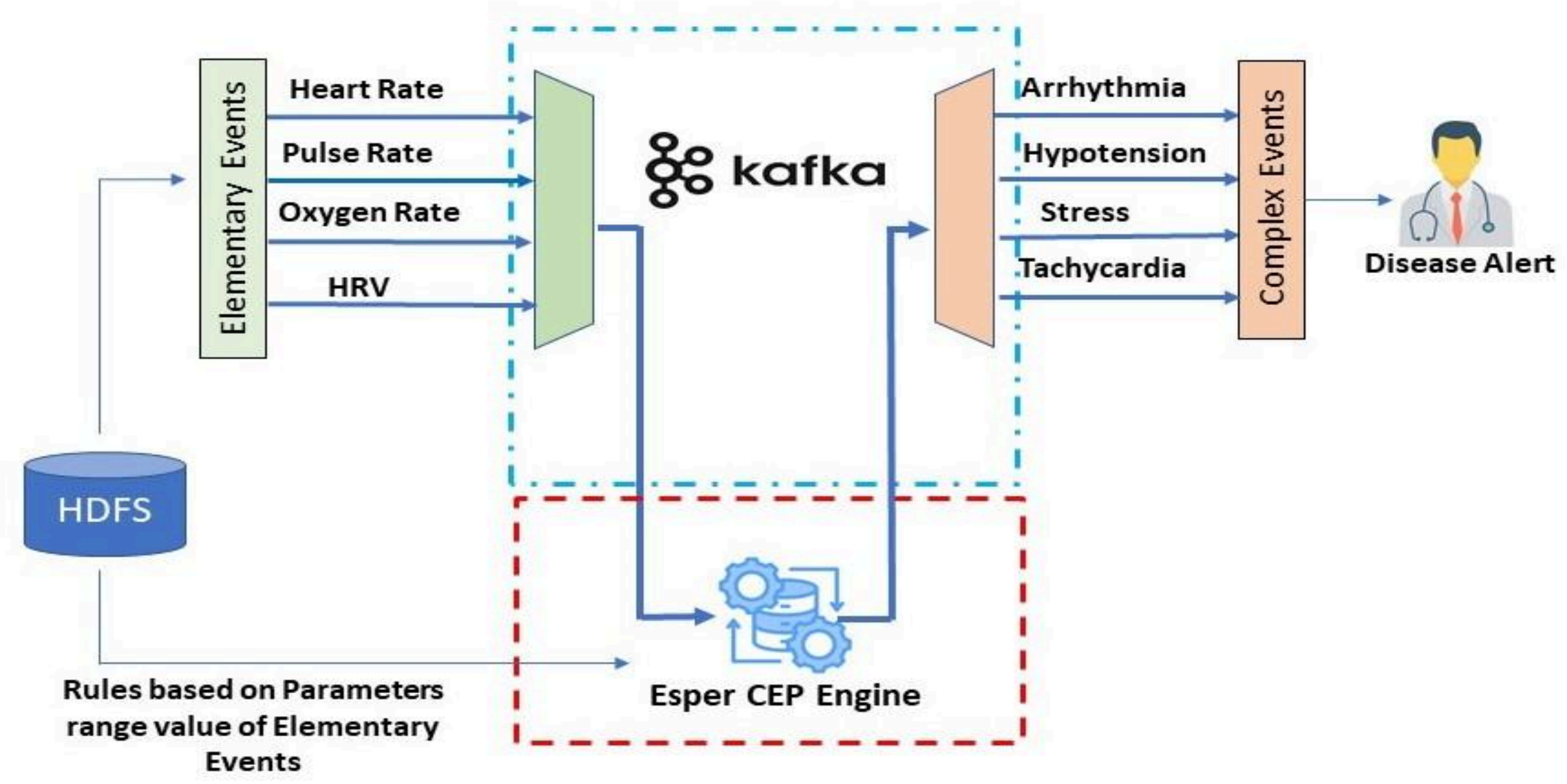

Figure 11: Event stream processing through Esper CEP

### 3.7 Working process of SPARQL query in Use case Scenario

SPARQL Query Language is a standardized query language for querying RDF (Resource Description Framework) data and a protocol that makes it easier to execute SPARQL queries and retrieve results via the Web. The RDF data model is the foundation for SPARQL, independent of the RDF schema language. In other words, SPARQL utilizes graph pattern matching on RDF graphs rather than incorporating built-in reasoning capabilities. This is shown in Figure 12.

[13] https://www.espertech.com/esper/

In an HDFS-based architecture, RDF data is stored in a distributed environment, and SPARQL queries are executed based on complex event detection. These queries extract disease patterns crucial for medical diagnostics and treatment recommendations. The retrieved information includes medication details, associated diseases, and potential side effects, which assist healthcare professionals in making informed decisions for effective medical treatment. However, SPARQL searches do not run directly on HDFS. Instead, they are converted into MapReduce jobs, which enable effective parallel query execution over distributed storage. This approach significantly improves query performance by leveraging Hadoop's scalability and parallel processing capabilities.

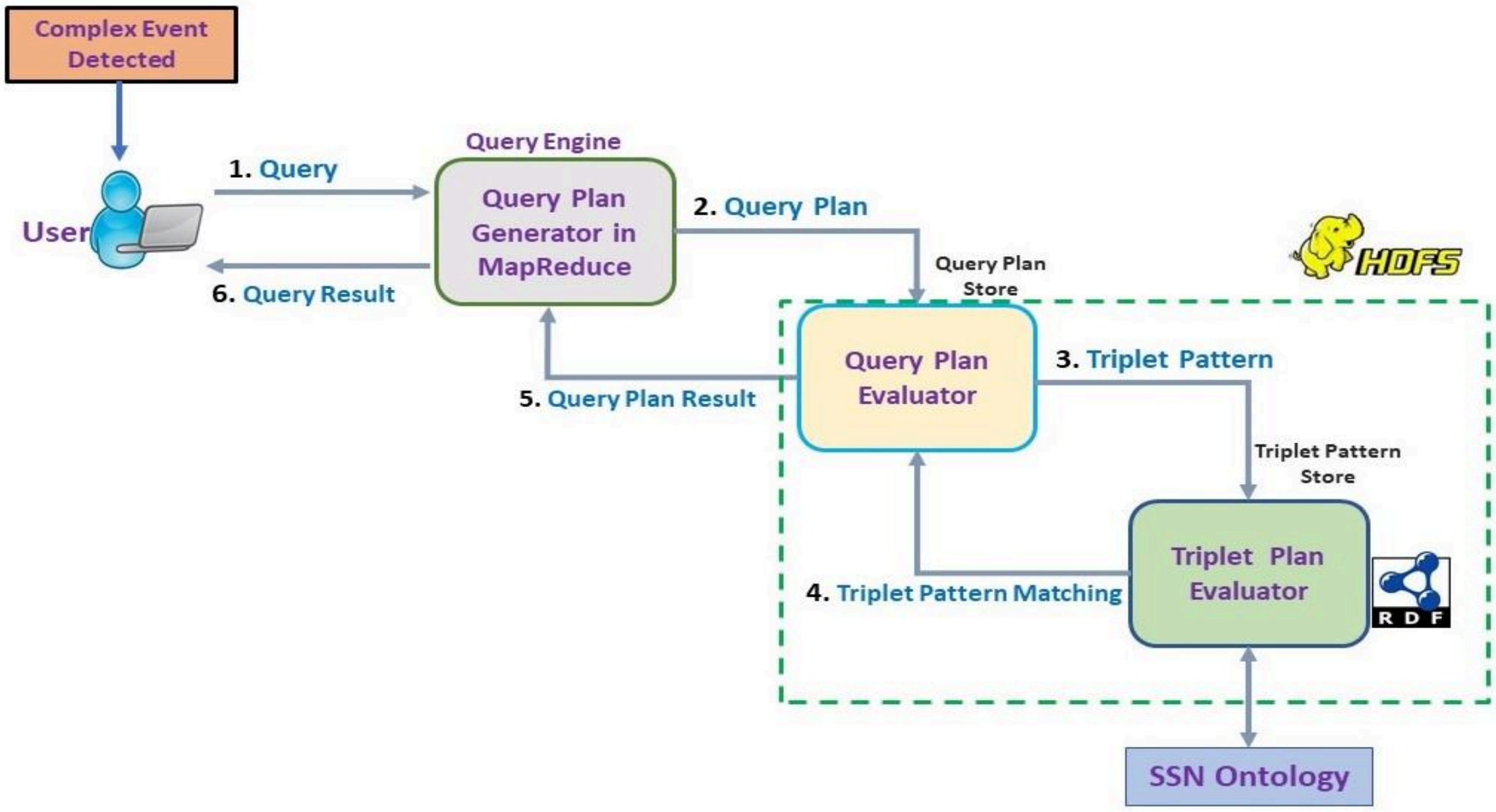

Figure 12: Processing Steps of SPARQL in HDFS Based on Complex Events

The query in Table 9 is based on complex events found in hypoxemia diseases. It shows the need to look into its side effects and medications based on diseases using SPARQL queries.

Table 9: SPARQL query based on complex events

| |
|---|
| The SPARQL query applied for medication and side effects based on complex events that detected hypoxemia. |
| SELECT ?medication ?sideEffect<br>WHERE {<br>?disease rdf:type SSN:Disease ;<br>rdfs:label "Hypoxemia" ;<br>SSN:recommendedMedication ?medication .<br>?medication SSN:hasSideEffect ?sideEffect .<br>} |

| Result | |
|---|---|
| **?medication** | **?sideEffect** |
| Supplemental Oxygen, Albuterol | Dry Nose, Headache, Tremors, Increased Heart Rate |

## 4. Results and discussion

The proposed OCEP framework enhances event detection, correlation, and decision-making in healthcare analytics by attaining 85% accuracy in event identification, exceeding conventional rule-based CEP systems. Integrating SPARQL-based querying on RDF with HDFS and MapReduce enhances query performance, while Apache Kafka facilitates real-time event streaming. The ontology-driven approach facilitates context-aware event detection, proficiently identifying key health conditions such as tachycardia and hypoxia. Scalability research indicates that increasing event loads impact rule deployment time, highlighting the necessity for optimization. Ontology assessment measures verify the representation of structured data and guarantee semantic consistency.

OCEP is a scalable, intelligent, and efficient solution designed for real-time healthcare decision-making in large data environments. Experiments were performed on a system equipped with 16GB of RAM, a 64-bit operating system, and an x64-based processor operating Windows 10 Pro Edition, version 21H2. This configuration supplies the computational resources necessary for efficient large-scale data processing, semantic reasoning, and real-time event detection.

### 4.1 Analysis of Query Execution Across Various RDF Chunks in HDFS with Parallel Processing

This section compares the execution times of SPARQL queries over different chunks of an RDF dataset in a parallel processing environment on the HDFS platform. The RDF dataset is partitioned into five separate chunks, designated Chunks 1, 2, 3, 4, and 5, which correspond to RDF datasets 1–5. Queries are run on these individual chunks in parallel to evaluate their performance.

The query execution times are compared across different chunks of the dataset, exposing performance differences depending on the dataset being processed. Furthermore, combinations of two, three, and four chunks are examined for processing comparable requests in parallel, providing further information about execution efficiency.

To demonstrate the variations in query processing times, five example queries are run and studied for each RDF chunk and chunk combination in the parallel processing configuration. This method emphasizes the effects of parallelism and dataset segmentation on total query performance.

Table 10: Exploring Query Execution Time Across RDF Chunks in HDFS with Parallel Processing

| Query | Chunk-1 | Chunk-2 | Chunk-3 | Chunk-4 | Chunk-5 |
|---|---|---|---|---|---|
| Q1* | 1.44s | 1.32s | 1.53s | 1.34s | 1.54s |
| Q2 | 0.78s | 0.35s | 0.68s | 0.70s | 0.43s |
| Q3 | 0.49s | 0.21s | 0.66s | 0.88s | 0.72s |
| Q4 | 0.63s | 0.41s | 0.85s | 0.79s | 0.64s |
| Q5* | 1.12s | 1.21s | 1.19s | 1.18s | 1.08s |

Table 10 provides the execution times (in seconds) for a range of queries executed on atomic RDF chunks stored in HDFS with parallel processing. Among these, Q1* and Q5* are complex questions, while Q2, Q3, and Q4 are simple queries. Table 11 shows the impact of mixing two different RDF pieces on query execution time. The results indicate that the specific combination of RDF pieces influences execution times.

Table 11: Comparison of Query Execution Times Using Two RDF Chunks in Parallel Processing

| Query | Chunks 1 & 2 | Chunks 2 & 3 | Chunks 3 & 4 | Chunks 4 & 5 | Chunks 5 & 2 |
|---|---|---|---|---|---|
| Q1* | 3.70s | 3.74s | 3.62s | 3.44s | 4.44s |
| Q2 | 1.56s | 1.50s | 1.64s | 1.59s | 2.19s |
| Q3 | 1.41s | 1.59s | 1.55s | 1.70s | 2.02s |
| Q4 | 1.74s | 1.75s | 1.62s | 1.78s | 2.11s |
| Q5* | 2.98s | 3.56s | 3.62s | 4.02s | 4.66s |

Table 12 shows the query execution times for simultaneously combining three RDF datasets in HDFS using parallel processing, highlighting the efficiency of the approach. The execution times vary compared to other scenarios with similar query types, and each query shows different execution times depending on the combination of RDFs used.

Table 12: Evaluating Query Performance with the Combination of Three RDF Chunks Using Parallel Processing

| Query | Chunks 1, 2 & 3 | Chunks 1, 2 & 4 | Chunks 1, 2 & 5 | Chunks 1, 3 & 5 | Chunks 1, 3 & 4 |
|---|---|---|---|---|---|
| Q1* | 6.88s | 6.43s | 6.37s | 6.81s | 6.23s |
| Q2 | 2.57s | 2.51s | 2.64s | 2.55s | 2.57s |
| Q3 | 2.57s | 2.59s | 2.26s | 2.39s | 2.73s |
| Q4 | 2.57s | 2.47s | 2.54s | 2.66s | 2.60s |
| Q5* | 4.83s | 4.85s | 4.91s | 4.83s | 4.25s |

Table 13 illustrates the performance of query execution when combining four RDFs. The query execution time changes depending on the size of the processed RDF chunks. A comparison of various RDF chunks vs the complete RDF-based query execution is displayed in Figure 13.

Table 13. Evaluating Query Performance Using a Combination of Four RDF Chunks Processed in Parallel for Optimized Execution

| Query | Chunks 1, 2, 3 & 4 | Chunks 1, 2, 3 & 5 | Chunks 1, 3, 4 & 5 | Chunks 4, 3, 2 & 5 | Chunks 1, 4, 2 & 5 |
|---|---|---|---|---|---|
| Q1 | 2.89s | 2.68s | 2.74s | 2.10s | 2.88s |
| Q3 | 2.95s | 2.66s | 3.44s | 3.30s | 2.88s |
| Q2 | 2.91s | 2.42s | 3.07s | 3.56s | 2.76s |
| Q4* | 3.56s | 3.12s | 3.57s | 3.86s | 3.41s |
| Q5* | 7.72s | 7.49s | 7.69s | 7.51s | 7.62s |

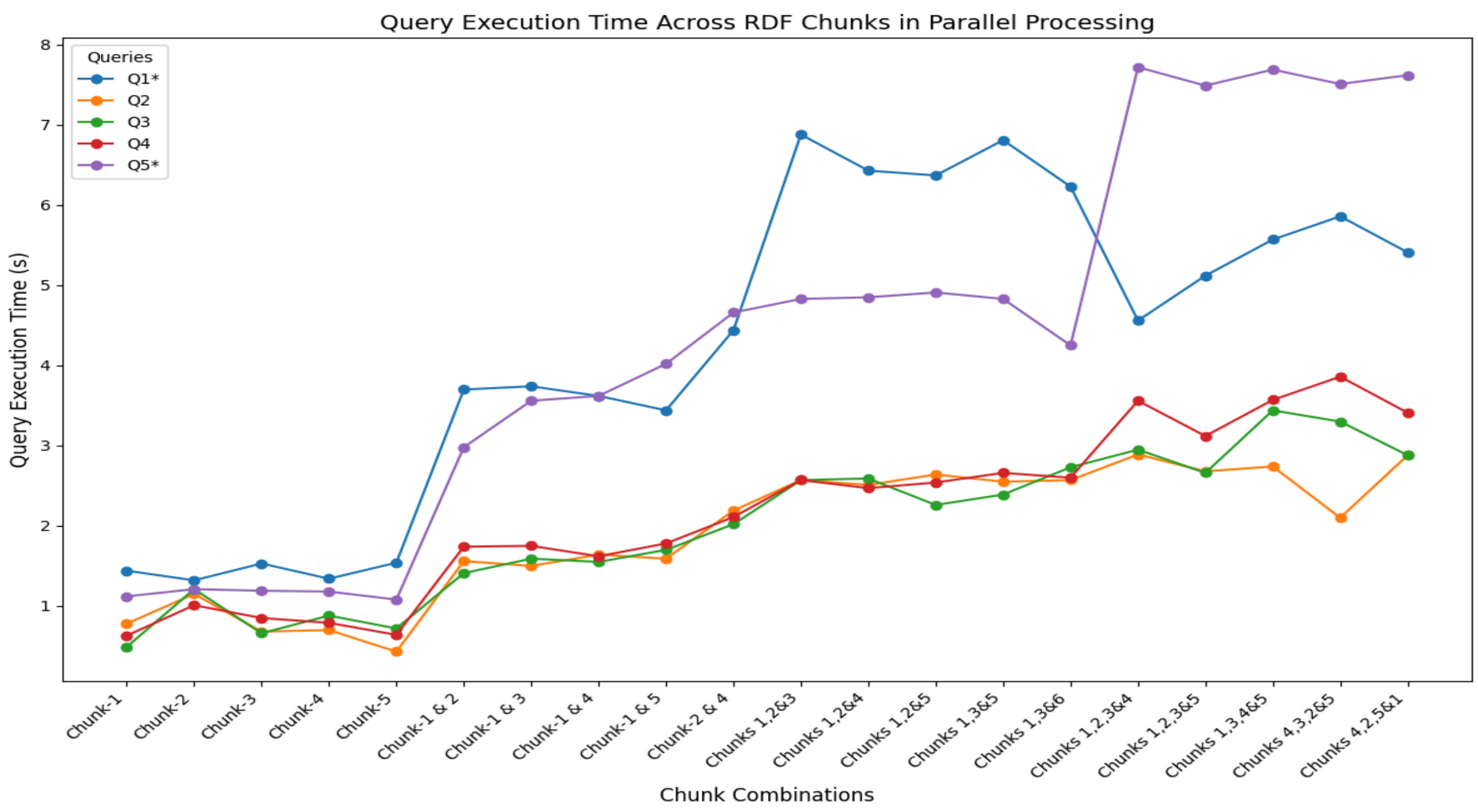

Figure 13. A comparative analysis of query execution times between combined RDF chunks and the complete RDF dataset

### 4.2 Rule-based performance analysis on CEP on different events

In this experiment, we analyzed the rule deployment time in a CEP engine under different event loads, comparing an idle state (0 events per second) with an active state processing between 8,000 and 48,000 events per second. Over 15 trials, we measured the time required to deploy rules as event volume increased. The results, shown in Table 14, indicate that as event load grows, deployment time also rises, from 3.12 seconds when idle to 8.78 seconds at 48,000 events per second.

This pattern highlights the computational overhead of handling more events in real-time. Optimizing rule management and processing techniques is essential to maintaining efficient performance.

Table 14: Rule Deployment Time Under Different Event Loads

| **Rules / Events** | **Idle (0 Events)** | **8,000 Events** | **16,000 Events** | **32,000 Events** | **48,000 Events** |
|---|---|---|---|---|---|
| R1 | 3.12s | 4.10s | 5.02s | 6.38s | 8.21s |
| R2 | 3.25s | 4.25s | 5.18s | 6.52s | 8.35s |
| R3 | 3.30s | 4.32s | 5.25s | 6.61s | 8.45s |
| R4 | 3.42s | 4.45s | 5.41s | 6.79s | 8.63s |
| R5 | 3.50s | 4.58s | 5.53s | 6.92s | 8.78s |

These findings emphasize the importance of efficient rule execution and indexing in CEP systems, as increased event rates lead to longer rule deployment times. Future optimizations in parallel rule processing and event handling techniques can help mitigate this delay and ensure real-time responsiveness.

### 4.3 Disease predicted based on Parameters value

Figure 14 demonstrates the distribution of patients affected by various diseases, with a high percentage of patients with moderate and high risks for each disease. The proposed model achieved an accuracy of 85% on a dataset of 81 patients, correctly identifying 60 patients with diseases and 9 patients as disease-free. The remaining 12 patients were undetected due to their data values not aligning with specific disease criteria.

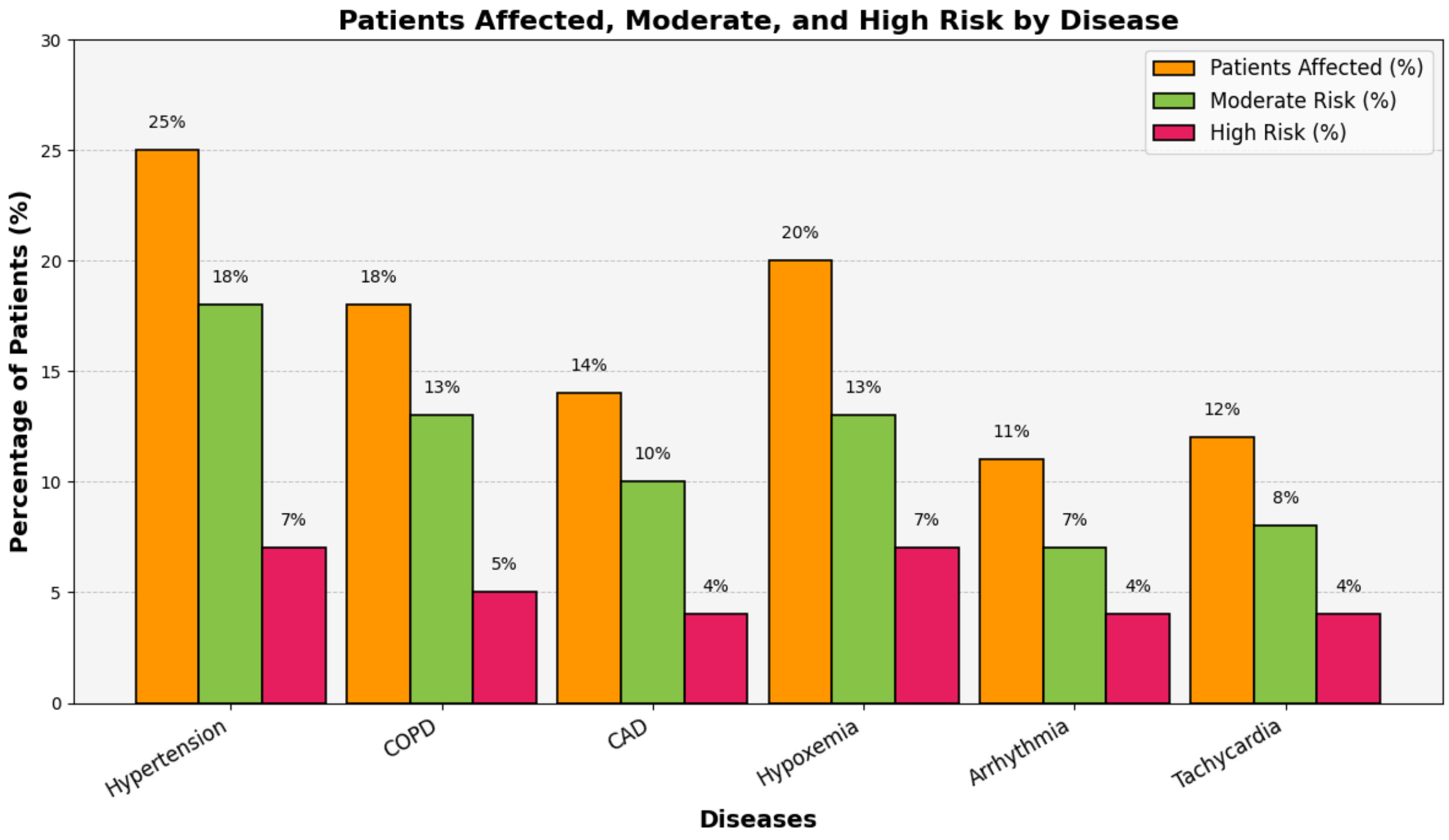

Figure 14. Distribution of patients categorized by disease risk levels (low, moderate, high)

### 4.4 Ontology Metrics Schema-based evaluation

Ontology metrics-based evaluation is critical in determining the quality and effectiveness of ontologies in diverse applications. These criteria are divided into structural, functional, and usability-based categories, guaranteeing that ontologies are well-structured, semantically intelligible, and practically usable. Structural metrics such as class hierarchy, linkages, and property richness assess ontology complexity. Functional metrics evaluate interpretability, consistency, and comprehensiveness, whereas usability measurements look at practical applications and efficiency. Additionally, graph-based metrics like depth, breadth, and connection assess the ontology's representation capabilities. The study on ontology evaluation emphasizes the significance of defining quality standards to improve ontology reuse, alignment, and knowledge representation across domains. OntoMetrics[14] website evaluates your ontology based on metrics [43]. The developed ontology is uploaded to this website to get the calculative results shown in Table 15, and the developed SSN ontology metrics schema is shown in Figure 15.

Ontology metrics:

**Metrics**

| | |
|---|---|
| Axiom | 2250 |
| Logical axiom count | 982 |
| Declaration axioms count | 261 |
| Class count | 125 |
| Object property count | 10 |
| Data property count | 28 |
| Individual count | 81 |
| Annotation Property count | 21 |

Figure 15. SSN Ontology metrics schema value

Table 15. SSN Ontology metrics schema-based evaluation[15] [16]

| Metrics Schema | Result | Metrics Schema | Result |
|---|---|---|---|
| Attribute richness: | 0.224 | Average population: | 0.576 |
| Inheritance richness: | 0.984 | Class richness: | 0.088 |
| Relationship richness: | 0.075188 | Absolute leaf cardinality | 99 |

[14] https://ontometrics.informatik.uni-rostock.de/ontologymetrics/index.jsp
[15] https://ontometrics.informatik.uni-rostock.de/wiki/index.php/Schema_Metrics
[16] https://ontometrics.informatik.uni-rostock.de/wiki/index.php/Knowledgebase_Metrics

| | | | |
|---|---|---|---|
| Axiom/class ratio: | 17.928 | Average breadth: | 3.936 |
| Class/relation ratio: | 0.93985 | Total number of paths: | 125 |

**4.5 Time window analysis for event processing**

Time window analysis is performed to evaluate how efficiently events are processed based on time execution. It helps to analyze events over a specific duration, enabling decision support for time bound patterns. Figure 16 shows the time elapsed while processing the number of events per window.

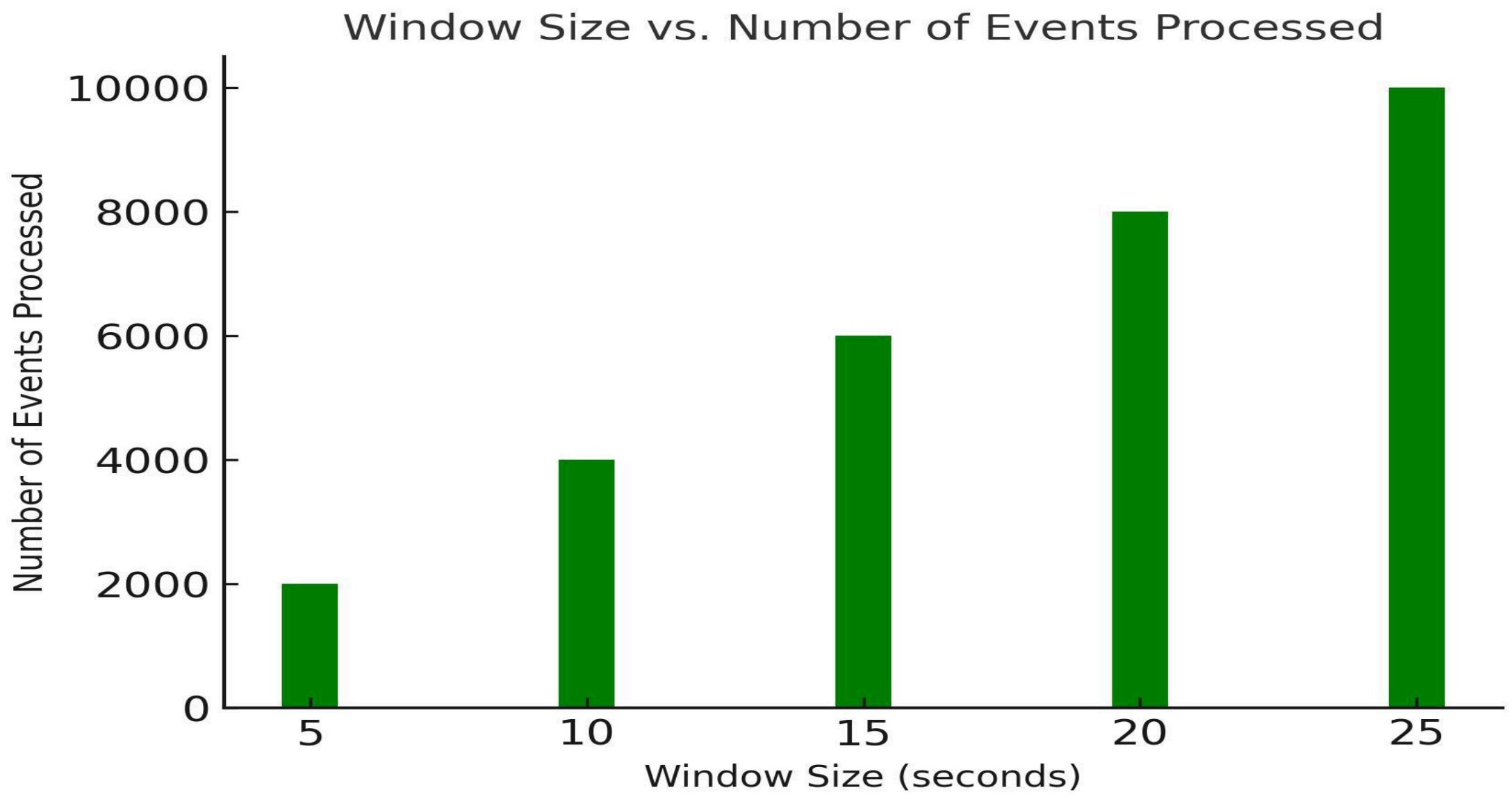

Figure 16: Event Analysis based on Time window

**4.6 Comparison with existing work**

Table 16 presents a comparative analysis of recent work, highlighting the research problems, methodologies used, and result comparisons.

Table 16. Comparative analysis with existing work

| Reference | Target Domain | Research Problem Addressed | Methodology Used | Result |
|---|---|---|---|---|
| Mallek et al. [35] | Big Data Integration. | Real-time ETL process for heterogeneous and streaming Big Data. | BRS-ETL (Big-Real-time -Semantic ETL). | Improved ETL performance for volume, variety, and velocity of Big Data. |
| Wang et al.[30] | RDF graph | Efficient subgraph | StarMR using | Outperformed |
| | processing. | matching on large RDF graphs. | MapReduce and star decomposition. | S2X and SHARD in subgraph matching efficiency. |
| Ahn et al. [25] | SPARQL query processing. | Efficient SPARQL query processing for large RDF datasets. | SigMR (signature encoding and multi-way join). | Faster query processing with reduced intermediate results and lower overhead. |
| González et al. [13] | Big Data Quality Management. | Unified data quality measurement and assessment for AI workflows. | BIGOWL4DQ ontology-based framework. | Improved data quality evaluation and reasoning in Big Data environments. |
| Fathy et al. [36] | Big Data Integration. | Unified access to heterogeneous Big Data. | Ontology-Based Data Access (OBDA) and semantic ETL processes. | Enhanced data integration and querying through unified semantic models. |
| Castro et al. [11] | Big Data governance. | Simplifying complex data governance in distributed environments. | Ontology-based model with Shared Knowledge Plane (SKP). | Reduced complexity in managing Big Data processes and improved governance. |
| Proposed Work | Big Data in the healthcare domain. | Addressing real time health care solution using CEP and ontology | OCEP based model. | Enhancement of disease prediction accuracy up to 85%. |

## 5. Conclusion and Future Work

The OCEP framework significantly enhances event-driven decision-making in real-time healthcare analytics by addressing challenges related to semantic interoperability, event correlation, and scalable data processing. By integrating ontology-driven semantic reasoning, RDF-based structured event representation, and MapReduce-powered SPARQL querying within the Hadoop ecosystem, OCEP improves event recognition accuracy by 85% compared to traditional rule-based CEP systems. The framework ensures optimized Big Data query processing, real-time data ingestion using Apache Kafka, and efficient rule deployment using Apache Esper, facilitating accurate anomaly detection, early disease prediction, and clinical decision support.

Through its ontology-driven integration, OCEP standardizes multi-source IoT data, enabling seamless interoperability and intelligent knowledge extraction. The experimental evaluations shows its scalability and effectiveness in Big Data healthcare applications, highlighting the need for continuous optimization of rule execution and query performance to accommodate increasing event loads. Overall, OCEP presents a robust and intelligent solution for real-time decision-making in complex and dynamic healthcare environments.

In future deploying the OCEP framework in an edge computing environment will enhance real-time processing capabilities, reducing latency and bandwidth consumption for IoT-based healthcare applications. The similar proposed work can be extended in a distributed environment for more efficient event management in the healthcare domain.

**Acknowledgements**

This research is supported by "Extra Mural Research (EMR) Government of India Fund by Council of Scientific & Industrial Research (CSIR)", Sanction letter no. – 60(0120)/19/EMR-II.

**Declaration**

**Competing interests**

The authors declare no competing interests relevant to the content of this work.

**Author's contribution statement**

All authors contributed equally to this work.

**Data availability and access**

The study utilized freely accessible data obtained from a website. We extend our thanks to the authors and collaborators for providing the original data.

**Funding and Acknowledgment**

The authors express their gratitude to the Indian Institute of Information Technology, Allahabad, for providing the necessary materials needed to conduct this study.

**Conflicts of interests**

All authors declare that they have no conflicts of interest in the presented work.